\input{aipcheck}

\documentclass[
    final            
  ,numberedheadings 
  ]
  {aipproc}
\usepackage{amssymb}
\newcommand{\md}{{\mathrm d}}
\newcommand{\sgn}{{\mathrm{sgn}}}
\newcommand{\lP}{\ell_{\rm P}}

\layoutstyle{8x11single}

\begin{document}

\title{Singularities and Quantum Gravity\footnote{Preprint Nos.\ IGPG--07/2--4, NSF--KITP--07--19}}


\author{Martin Bojowald}{
  address={Institute for Gravitational Physics and Geometry, The Pennsylvania State University,\\ 104 Davey Lab, University Park, PA 16801, USA\\[3mm]
Kavli Institute for Theoretical Physics, University of California, Santa Barbara, CA 93106, USA}
}

\begin{abstract}
  Although there is general agreement that a removal of classical
  gravitational singularities is not only a crucial conceptual test of
  any approach to quantum gravity but also a prerequisite for any
  fundamental theory, the precise criteria for non-singular behavior
  are often unclear or controversial. Often, only special types of
  singularities such as the curvature singularities found in isotropic
  cosmological models are discussed and it is far from clear what this
  implies for the very general singularities that arise according to
  the singularity theorems of general relativity. In these lectures we
  present an overview of the current status of singularities in
  classical and quantum gravity, starting with a review and
  interpretation of the classical singularity theorems.  This suggests
  possible routes for quantum gravity to evade the devastating
  conclusion of the theorems by different means, including modified
  dynamics or modified geometrical structures underlying quantum
  gravity. The latter is most clearly present in canonical
  quantizations which are discussed in more detail. Finally, the
  results are used to propose a general scheme of singularity removal,
  quantum hyperbolicity, to show cases where it is realized and to
  derive intuitive semiclassical pictures of cosmological bounces.
\end{abstract}

\maketitle

\section{Overview}

Physical theories are always idealizations without which the
complexity of nature would be too great to fathom. Theoretical physics
is, mostly very successfully, based on assumptions needed to formulate
equations, find solutions and use them to describe, explain and
further investigate physical phenomena.

Sometimes, however, these assumptions may not be general enough for
all purposes. When they are violated, the theory breaks down which
mathematically appears as the development of singularities.  An
example is given by the use of continuous fields rather than discrete
atomic structures in condensed matter physics. When fields vary too
strongly on small length scales, such as in shock waves, singularities
can occur in continuous field equations even though the basic,
discrete physical description remains valid.  Usually, deviations
between solutions and observations increase before a mathematical
singularity is reached. It is then clear that the approximate
description can no longer be trusted beyond a certain point. But
observations are not always available in such regimes where
singularities are approached and an interpretation of mathematical
singularities becomes more difficult.  This is the case especially for
gravity where observations of strong field regimes are lacking.

Singularities in general relativity therefore play a special and dual
role. First, the classical importance of singularities can be
questioned since there are always assumptions behind special solutions
or general theorems leading to singularities. But classical
singularities in general relativity also provide an excellent chance
to derive implications for the structure of space-time described by
general relativity. When the theory breaks down, lessons for
space-time structure result which can be especially important for the
development of quantum gravity possibly replacing general relativity
around classical singularities.

We will first review the classical singularity theorems and sketch
their main idea of proof. This will allow us to see which assumptions
enter the theorems and what their main conclusions are. These
theorems, rather than special singular solutions, define the
singularity problem of general relativity. Their statements provide
the measuring rod which any proposal for singularity resolution has to
be compared with.

The following section will deal with potential examples for
singularity resolution which have been proposed in quantum gravity,
mostly in string theory and in canonical quantizations. The examples
are not intended to be complete but to indicate the general types of
ideas that have been put forward. (See also \cite{KITPSumm} for a
summary talk.) Here, {\em quantum hyperbolicity} will be formulated as
a general principle and it will be shown to require characteristic
properties of quantum gravity to be realized.

Our specific formulation of the principle is worded in the language of
canonical quantum gravity which is described in a subsequent
section. We start with an explanation of the difficulties encountered
in the first attempts of Wheeler--DeWitt quantizations, and show how
their resolution naturally leads to loop quantum gravity. Quantum
geometry and quantum dynamics in this framework are then discussed at
length, as they provide the main pillars for any attempt to address
the fate of classical singularities.

In this framework, loop quantum cosmology has led to explicit
constructions of dynamical laws from which non-singular behavior can
be derived in several models. This is where the principle of quantum
hyperbolicity is currently realized without counterexamples. Loop
quantum cosmology thus provides the most general scheme of singularity
removal available at present, and it can be used for explicit
scenarios.

In most cases, however, the basic description around a classical
singularity requires deep quantum regimes which do not lend themselves
easily to intuitive interpretations. It can thus be helpful to
develop effective descriptions which capture some quantum effects but
are otherwise based on classical concepts. This is available in
semiclassical bounce pictures which provide examples of how
singularities can be avoided through bounces in certain regimes. It
also provides the basis for perturbation theory to compute
phenomenological and potentially observable effects of metric modes
and other fields traveling through a classical singularity.

Although ideas in all five lectures are closely related, the text of
any section can be read largely independently of the others.

\begin{figure}
  \includegraphics[height=.1\textheight]{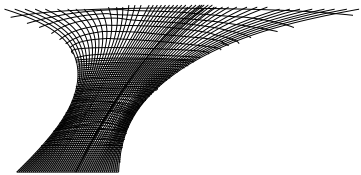}
  \caption{Illustration of a smooth manifold dissolving into a
    classical singularity, or into a discrete space where quantum
    gravity takes over.\label{SurfaceSing}}
\end{figure}

\section{Classical Singularities}

General relativity\footnote{We will mainly follow the notation used in
  \cite{Wald}. In particular, we adopt the abstract index notation
  where objects such as $T_{ab}$ denote tensors rather than single
  components in specific coordinates. Expressions only true in certain
  coordinates are indicated with greek indices. We do not use
  different types of indices for space-time tensors and spatial
  tensors; it is rather clear from the context whether an object
  refers to a space-time or spatial manifold. Note also that $N$ and
  $N^a$ denote different objects, a function in the first and a vector
  field in the second case, although traditionally the same letter is
  used.} describes the gravitational field as a consequence of the
space-time structure determined in terms of the metric tensor as a
solution to Einstein's equation $G_{ab}=8\pi G T_{ab}$ sourced by the
energy-momentum tensor $T_{ab}$ coupled through the gravitational
constant $G$.  In component form, these are coupled, non-linear
partial differential equations of second order for the space-time
metric $g_{ab}$. Unlike other fundamental field theories, they are
generally covariant under arbitrary changes of space-time coordinates
which shows that there is no background space and time on which fields
are defined.  This implies gauge symmetries for the space-time metric.

Gauge theories in general are more systematically analyzed in a
canonical formulation where one uses fields and their momenta rather
than fields and their time derivatives. One is thus breaking up
space-time tensors into spatial and time components which hides the
underlying general covariance. To introduce this in general
relativity, one foliates space-time into a family of spatial slices
$\Sigma_t$ parameterized by an arbitrary time coordinate $t$. The
space-time metric adapted to this foliation can be written as
\begin{equation} \label{CanonMetric}
\md s^2=-N^2\md t^2+q_{ab}(\md x^a+ N^a \md t) (\md x^b+N^b\md t)
\end{equation}
with the {\em spatial metric} $q_{ab}$ on $\Sigma_t$, the {\em lapse
  function} $N$ and the spatial {\em shift vector} $N^a$. The
components are all functions of spatial coordinates on $\Sigma_t$ as
well as time $t$.  When inserted in Einstein's equation, evolution
equations in $t$,
\begin{equation} \label{qdot}
 \dot{q}_{ab}=\frac{16\pi GN}{\sqrt{\det q}} (2p_{ab}-p^c_c q_{ab})+
 2D_{(a}N_{b)}
\end{equation}
and
\begin{eqnarray} \label{pdot}
 \dot{p}_{ab} &=& -\frac{N\sqrt{\det q}}{16\pi G} ({}^{(3)}\!R^{ab}-
 \frac{1}{2}{}^{(3)}\!R q^{ab})
 + \frac{8\pi GN}{\sqrt{\det q}}
 q^{ab}(p^{cd}p_{cd}-\frac{1}{2}(p^c_c)^2)- \frac{32\pi GN}{\sqrt{\det
     q}}(p^{ac}p_c{}^b-\frac{1}{2}p^{ab}p^c_c)\\
 &&+ \frac{\sqrt{\det q}}{16\pi G}(D^aD^bN-q^{ab}D_cD^cN)
 + \sqrt{q} D_c\frac{p^{ab}N^c}{\sqrt{\det q}}-
 2p^{c(a}D_cN^{b)}\nonumber\,,
\end{eqnarray}
result from the space-space components written in first order form for
the spatial metric $q_{ab}$ and its momentum
$p^{ab}=\sqrt{q}(K^{ab}-K^c_cq^{ab})/16\pi G$ which is related to
extrinsic curvature $K_{ab}=(2N)^{-1}(\dot{q}_{ab}-D_aN_b-D_bN_a)$ of
the spatial slices in space-time. In addition, there are constraints
\begin{equation} \label{constr}
 \frac{\sqrt{\det q}}{16\pi G}{}^{(3)}\!R- \frac{16\pi
G}{\sqrt{\det q}} (p_{ab}p^{ab}-{\textstyle\frac{1}{2}}(p^a_a)^2)=0
\quad,\quad D_bp_a^b=0
\end{equation}
resulting from the time-time and time-space components of Einstein's
equation in vacuum. In all equations, $D_a$ denotes the spatial covariant
derivative compatible with $q_{ab}$, and ${}^{(3)}\!R_{ab}$ its Ricci
tensor. It is clear from those equations that the components $N$ and
$N^a$ do occur in the evolution equations of $q_{ab}$, but are
themselves unrestricted (except by the condition that
(\ref{CanonMetric}) must be a Lorentzian metric).

\subsection{Initial value problem}

We thus obtain an initial value problem only once lapse function $N$
and shift vector $N^a$ have been specified throughout space-time as a
gauge choice. (See \cite{CauchyProblem} for a discussion of the types
of initial value problems realized in general relativity.) Their
equations can be interpreted as determining the manifold structure of
space-time, which is most clearly seen when using the gauge source
function $\Gamma^{\tilde\mu}=
g^{\nu\lambda}\Gamma^{\tilde\mu}_{\nu\lambda}$.  When conditions are
imposed by fixing $\Gamma^{\tilde\mu}$, a gauge is determined. Notice
that the Christoffel symbol $\Gamma^{\tilde\mu}_{\nu\lambda}$ does not
form a tensorial object, which we mark here by using the tilde on one
index, and thus fixing its values even to zero restricts the choice of
coordinate systems.

Thus, fixing $\Gamma^{\tilde{\mu}}$ implies conditions on the gauge
functions $N$ and $N^a$.  Through the usual relation $
\Gamma^{\mu}_{\nu\lambda}= \frac{1}{2}g^{\mu\rho}
(\partial_{\nu}g_{\lambda\rho}+ \partial_{\lambda}g_{\nu\rho}-
 \partial_{\rho}g_{\nu\lambda})$
between $\Gamma^{\mu}_{\nu\lambda}$ and $g_{ab}$, one has, for instance,
\[
 \partial_tN-N^a\partial_aN=N^2(K^a_a-n^a\Gamma_a)
\]
which is an identity between $\Gamma^{\tilde{\mu}}$ and the canonical
metric components if $\Gamma^{\tilde{\mu}}$ is kept free. (The vector
field $n^a$ is the unit normal vector to the spatial slices.) If
$\Gamma^{\tilde{\mu}}$ is prescribed as a gauge choice, however, the
equation becomes an evolution equation for $N$ in this chosen gauge

Moreover, we have
\begin{equation} \label{GammaCoord}
 \Gamma^{\tilde\nu}= g^{\mu\lambda}\Gamma^{\rho}_{\mu\lambda}
 \delta^{\tilde\nu}_{\rho}= -g^{\mu\lambda}
 \nabla_{\mu}\delta^{\tilde\nu}_{\lambda}= 
-\nabla_{\mu}\nabla^{\mu}x^{\tilde\nu}
\end{equation}
such that prescribing $\Gamma^{\tilde\mu}$ poses ``evolution''
equations for space-time coordinates $x^{\tilde\mu}$. A common choice
is the harmonic gauge $\Gamma^{\tilde\mu}=0$ where coordinates are
harmonic functions. Once coordinates on an initial spatial slice are
chosen, space-time coordinates are determined as a solution of
(\ref{GammaCoord}) by fixing the gauge source function.

The space-time manifold as a topological set equipped with an atlas of
coordinate charts is thus, to some degree, part of the solution as a
consequence of general covariance.  This is entirely different from
other field theories for fields on a given background (metric)
manifold as they are used for the remaining fundamental forces. This
does not only give rise to complicated conceptual and technical issues
when a quantization is attempted, as we will see later, but also to
new physical features already present in the classical theory which
are sometimes disturbing. We are able to derive properties of space
and time themselves, and about their ends. Although not much is known
about general solutions of Einstein's non-linear partial differential
equations, there is one general feature common to most realistic
solutions of general relativity: Space-time cannot be extended
arbitrarily but develops boundaries where the classical theory breaks
down.

\subsection{Singularities}

Since it is difficult to determine solutions or even asymptotic
properties in general, a useful idea is to employ test particles as
probes of possible space-time boundaries. One thus studies how test
objects behave in a given solution to Einstein's equation and whether
their motion, as described by the classical theory, has to stop at a
certain point. If this occurs, the failure to move the test object
further can only be attributed to a boundary of space-time itself
since no interactions are included which could destroy the object.
Quite surprisingly, this procedure allows far-reaching conclusions
with only the slightest input from Einstein's equation
\cite{SingTheo,HawkingEllis}.

The precise criterion for a space-time singularity in this sense is
geodesic incompleteness: space-time is singular if a geodesic, i.e.\ a
word-line of a freely falling test object, exists which is not complete
and not extendible. Thus, a curve $e\colon {\cal I}\to M$ exists whose
tangent vector satisfies $\dot{e}^a\nabla_a\dot{e}^b=0$, i.e.\ it is a
piece of an affinely parameterized geodesic defined on a proper subset
${\cal I}\subset{\mathbb R}$ which cannot be extended to be defined on
a larger subset ${\cal J}\supset{\cal I}$. As the trajectories of
freely falling particles, geodesics describe test objects subject only
to the gravitational force.

Before we sketch the main proof of a singularity theorem, we collect
the typical assumptions and properties used:
\begin{itemize}
 \item Positive energy conditions, translated to positive curvature
   through Einstein's equation, imply
 focusing of families of geodesics and thus self-intersections and caustics.
\item Topological properties of space-time such as global
  hyperbolicity or spatial non-compactness then allow one to relate
  properties of geodesic families to space-time properties.
\item Appropriate initial configurations select physical situations,
  such as an everywhere expanding/contracting spatial slice
  (cosmology) or the existence of a trapped surface (black holes)
  where, with the preceding two assumptions, singularities are bound
  to occur.
\end{itemize}
The proof we sketch here demonstrates the importance of all these
assumptions and illustrates how they could potentially be
circumvented. To be specific, we focus on singularities as they occur
in black holes, requiring the existence of a trapped surface as an
initial condition. More details and different types of theorems can be
found in \cite{HawkingEllis,Senovilla}.

\subsubsection{Trapped surfaces}

\begin{figure}
  \includegraphics[height=.15\textheight]{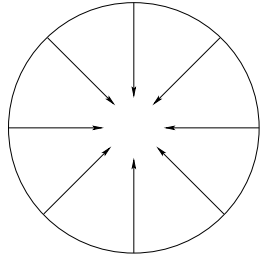}
  \caption{The ingoing normal family of a sphere is contracting
    because its cross-section area decreases along the affine
    parameter of the geodesic family.\label{Ingoing}}
\end{figure}

A trapped surface is a compact, 2-dimensional smooth space-like
submanifold $T\subset M$ such that the families of outgoing as well as
ingoing future-pointing null normal geodesics are contracting
\cite{TrappedSurface}. A geodesic family is defined by specifying a
transversal vector field on a submanifold which uniquely determines a
family of geodesic curves through each point of the submanifold in a
direction given by the vector field at that point.  For a compact
2-surface, we can use the inward pointing and outward pointing null
normals as those vector fields, defining the ingoing and outgoing null
geodesic families.  As shown in Fig.~\ref{Ingoing}, the ingoing family
of normal geodesics is usually contracting in the sense that its
cross-section area decreases, but conversely one intuitively expects
the outgoing family to be expanding. A trapped surface requires even
the outgoing family to be contracting and thus occurs only under
special circumstances.

To define expansion and contraction of families of null geodesics
formally, we use their tangent vector field $k^a$. Then, {\em
  expansion} is defined as $\theta:=\nabla_ak^a$. (This is similar to
non-relativistic fluid dynamics where the divergence of a velocity
field gives the expansion of fluid volume elements.)  Similarly, one
can use the tensor $B_{ab}=\nabla_bk_a$ to introduce {\em shear}
$\sigma_{ab}$ as its symmetric, trace-free part and {\em rotation}
$\omega_{ab}$ as its anti-symmetric part (paying due attention to the
fact that $k^a$ is null in the precise definition which we are not
going to need here).  These tensors are subject to evolution equations
along the family, following from the geodesic equation:
\begin{equation} \label{GeodesicDev}
 k^c\nabla_cB_{ab} =
k^c(\nabla_b\nabla_ck_a+R_{cba}{}^dk_d)
= \nabla_b(k^c\nabla_ck_a)-
(\nabla_bk^c)(\nabla_ck_a)+ R_{cba}{}^d k^ck_d
= -B^{c}{}_bB_{ac}+
R_{cba}{}^d k^ck_d
\end{equation}
where we used the Leibniz rule, the geodesic equation and the relation
$(\nabla_a\nabla_b-\nabla_b\nabla_a)\omega_c=R_{abc}{}^d\omega_d$,
valid for any smooth dual vector field $\omega_c$, which introduces the
Riemann curvature tensor $R_{abc}{}^d$.

A simple example for the computation of expansion $\theta$ is given by
the light cone in Minkowski space whose generators define a family of
null geodesics.  The null tangent vector field is 
\[
 k^a={\rm sgn}(t)\left(\frac{\partial}{\partial r}\right)^a+
\left(\frac{\partial}{\partial t}\right)^a={\rm
  sgn}(t)\left(\frac{x}{r}\left(\frac{\partial}{\partial x}\right)^a+
  \frac{y}{r}\left(\frac{\partial}{\partial y}\right)^a+
  \frac{z}{r}\left(\frac{\partial}{\partial
      z}\right)^a\right)+\left(\frac{\partial}{\partial t}\right)
\]
and in Cartesian coordinates of Minkowski space we have
$B_{\mu\nu}=\partial_{\nu}k_{\mu}$ simply in terms of partial
derivatives.  Taking the trace, one obtains $\theta=2{\rm sgn}(t)/r$
which diverges at the tip of the cone where the null geodesics
intersect and form a caustic.

\begin{figure}
  \includegraphics[height=.2\textheight]{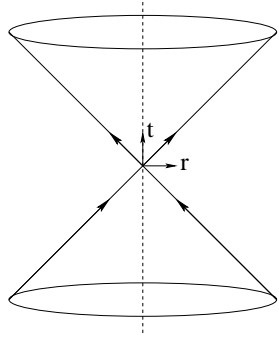}
  \caption{Light cone in Minkowski space.\label{LightCone}}
\end{figure}

The relation between self-intersections of geodesic families and the
divergence of their expansion is general: Take a geodesic family which
initially fills all of space-time, i.e.\ which emanates transversally
from a 3-dimensional submanifold. We can thus define three independent
vector fields $\kappa_i^a=(\partial/\partial \sigma_i)^a$ on a family
of geodesics parameterized by coordinates $\sigma_i$ of the
3-dimensional submanifold in addition to the null tangent
$k^a=(\partial/\partial t)^a$ along geodesics.  A caustic (or
conjugate point) forms if $\kappa_i^a$ becomes degenerate as a
$3\times3$ matrix since the infinitesimal separation between different
geodesics then vanishes in at least one direction. The matrix
$\kappa^a_i$ is related to expansion by
\[
 \kappa^b_iB_b^a= \kappa^b_i\nabla_bk^a= k^b\nabla_b\kappa^a_i
\]
using the commutation of coordinate derivatives,
$[k,\kappa_i]^a=k^b\nabla_b\kappa_i^a- \kappa_i^b\nabla_bk^a=0$.
Then indeed, expansion
\[
 \theta=B^a_a=(\kappa^{-1})^i_ak^b\nabla_b\kappa^a_i=
  k^b\nabla_b\log|\det \kappa^a_i|
\]
diverges when $\kappa^a_i$ becomes degenerate,\footnote{We have used
  $(\det\kappa^a_i)^{-1}\nabla_b\det\kappa^a_i=
  (\kappa^{-1})^i_a\nabla_b\kappa^a_i$ which follows easily from
  $\det\kappa^a_i=\frac{1}{4!}\epsilon_{abcd}\epsilon^{ijkl}
  \kappa^a_i \kappa^b_j \kappa^c_k \kappa^d_l$.} $\det\kappa^a_i=0$.

From the geodesic deviation equation (\ref{GeodesicDev}) we obtain
the {\em Raychaudhuri
equation} as its trace:
\begin{equation}
 \dot{\theta}=k^c\nabla_c\theta= g^{ab}k^c\nabla_cB_{ab}= 
-{\textstyle\frac{1}{2}}\theta^2-
\sigma_{ab}\sigma^{ab}+ \omega_{ab}\omega^{ab}-R_{ab}k^ak^b\,.
\end{equation}
Thus, if a family of geodesics is non-rotating,\footnote{For a
  geodesic congruence one can see, using the Frobenius theorem, that
  this is realized whenever the congruence is orthogonal to a
  hypersurface.} which means $\omega_{ab}=0$ and curvature is
non-negative as a consequence of energy conditions (such as the null
energy condition $T_{ab}k^ak^b\geq0$ for all null vectors $k^a$) and
Einstein's equation then $\theta$ always decreases along geodesics in
the family. This is the focusing effect of gravity and plays a major
role in deriving singularity theorems.

Quantitatively, we have $\dot{\theta}\leq-\frac{1}{2}\theta^2$ such
that after integration $\theta^{-1}\geq
\theta_0^{-1}+\frac{1}{2}(t-t_0)$ starting from initial values
$\theta_0$ at $t=t_0$. This is the place where initial values for
geodesic families enter: For negative $\theta_0$, $\theta$ must
diverge after finite time no larger than $t-t_0=2/|\theta_0|$.
Geodesics must intersect before that time and form a caustic (focal
point).

\subsubsection{From caustics to singularities}

Caustics are singularities where a geodesic family ceases to define a
smooth submanifold of space-time. But they are not physical
singularities since space-time itself is usually well-defined at
points where light rays intersect. To relate the occurrence of
caustics to space-time singularities we need one more ingredient
beyond the initial conditions and positive energy assumptions together
with Einstein's equation already used.

\begin{figure}
  \includegraphics[height=.15\textheight]{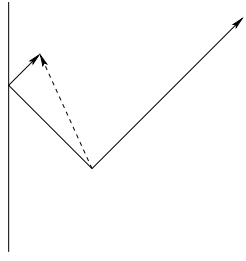}
  \caption{Causal diagram in spherical coordinates showing the ingoing
    null geodesic family of a 2-sphere centered at the origin of
    spherical coordinates. The family has a focal point at the origin
    (left vertical line). Any point beyond the the focal point can be
    reached by a time-like curve from the initial 2-sphere.
    \label{IngoingCausal}}
\end{figure}

The required basic statement from differential geometry is the
well-known property of geodesics as extremizing arc-length. Space-like
geodesics extremize arc-length
$\ell_e=\int_e\sqrt{g_{ab}\dot{e}^a\dot{e}^b}\md t$ among curves
between two given points and minimize it unless there is a focal point
between the two points. Similarly, time-like geodesics extremize
proper time $\tau_e=\int_e\sqrt{-g_{ab}\dot{e}^a\dot{e}^b}\md t$
between two points and maximize it unless there is a focal point
between the two points.  For a null geodesic such an extremization
condition is not possible since the norm of a null vector vanishes and
thus any null curve has zero length. But there is an analog to the
statement that minimization of arc-length by space-like geodesics or
maximization of proper time by time-like ones ceases beyond focal
points: Any point on a null geodesic beyond a focal point can be
reached by a time-like curve. In a causal diagram in spherical
coordinates, this is depicted in Fig.~\ref{IngoingCausal}.

\begin{figure}
  \includegraphics[height=.15\textheight]{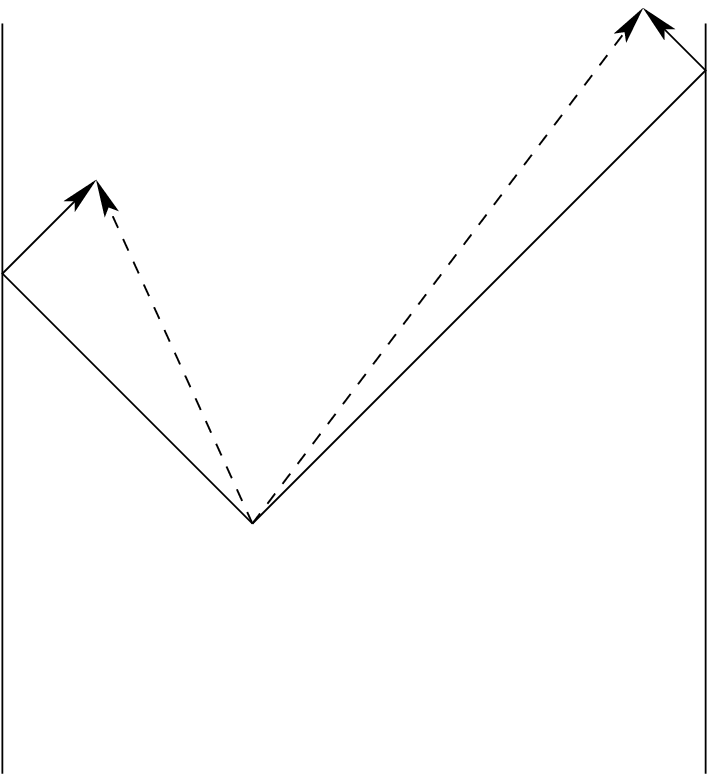} \hspace{.2\textwidth}
  \includegraphics[height=.15\textheight]{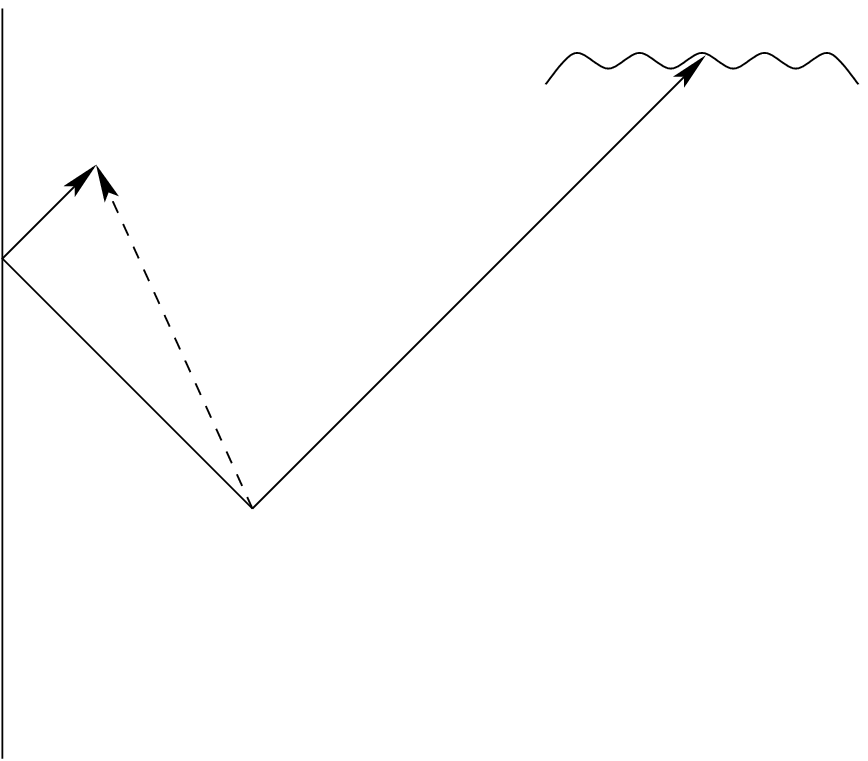}
  \caption{Causal diagrams of a spatially compact manifold with two
    focal points of null geodesic families (left) and of the outgoing
    null geodesic family encountering a space-time singularity in the
    spatially non-compact case (right).\label{CausalCompactSing}}
\end{figure}

According to the Raychaudhuri equation, together with the usual
assumptions of positive energy, every null normal geodesic family
starting from a trapped surface must develop a focal point since the
initial expansions are negative by definition of a trapped surface.
This is always realized for ingoing geodesics, but can it be possible
for outgoing ones? There is no difficulty if space is compact, as
shown in the left part of Fig.~\ref{CausalCompactSing}, since there will
be another coordinate center or a periodic identification encountered
by the outgoing family. In fact, for a compact spatial manifold one
cannot clearly distinguish between ingoing and outgoing null normals
of a 2-dimensional space-like submanifold. But for non-compact spatial
topology the outgoing null geodesics continue to go on forever and
can never be caught up with by a time-like geodesic.

There is thus a contradiction between the Raychaudhuri equation, in
the context of positive curvature and given the existence of a trapped
surface, and space-time topology for non-compact space. At this stage,
finally, topological conditions are needed and we are able to
translate caustics of geodesic families into space-time singularities.
The contradiction can only be avoided by concluding that incomplete
geodesics exists. The outgoing geodesic family does not develop a
focal point, despite of the Raychaudhuri equation, because it can
simply not be extended arbitrarily. When we formulated the
contradiction there was the hidden assumption that the null geodesics
in the family can be extended arbitrarily, i.e.\ we assumed them to be
complete. The only way to avoid the contradiction is to conclude that
incomplete null geodesics must exist: space-time is singular in the
sense of geodesic incompleteness (see the right part of
Fig.~\ref{CausalCompactSing}).

\subsubsection{Scheme of singularity theorems}

The proof sketched above illustrates the general assumptions and
conclusions used in singularity theorems:
\begin{enumerate}
\item {\em Initial conditions} ensure the existence of geodesic
  families with negative expansion.  Typical cases are trapped
  surfaces, implying black hole singularities, or spatial slices whose
  expansion or contraction is bounded away from zero everywhere,
  giving rise to cosmological singularities.
\item Using {\em positive energy conditions} together with 
  Einstein's
  equation in the Raychaudhuri equation implies focusing. With the
  initial conditions specified, caustics develop in finite time.
\item A caustic in general is only a ``singularity'' of the geodesic
  family, not of space-time. A singularity theorem finally
  results together with {\em topological assumptions} rendering a
  caustic into an obstruction to geodesic completeness.
\end{enumerate}
An important consequence is that dynamics of the gravitational field
is not used very specifically, but only to translate positive energy
conditions into focusing.

\subsubsection{Example: Schwarzschild geometry}

\begin{figure}
 \includegraphics[width=.15\textwidth]{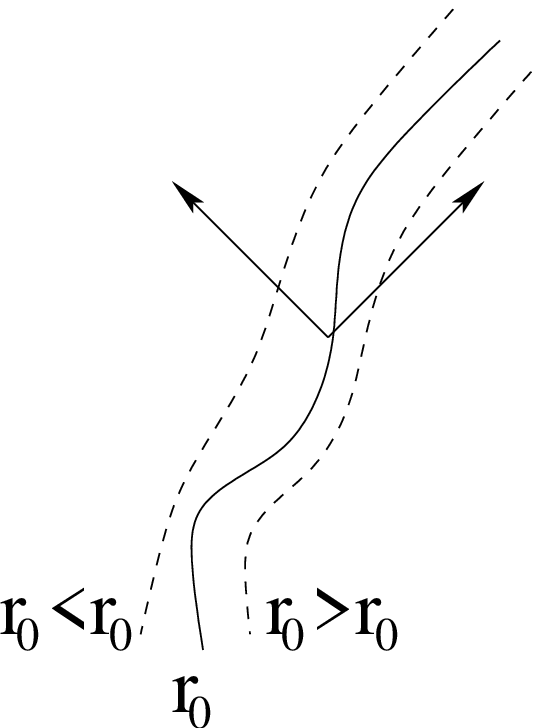}
 \hspace{.1\textwidth}
  \includegraphics[width=.2\textwidth]{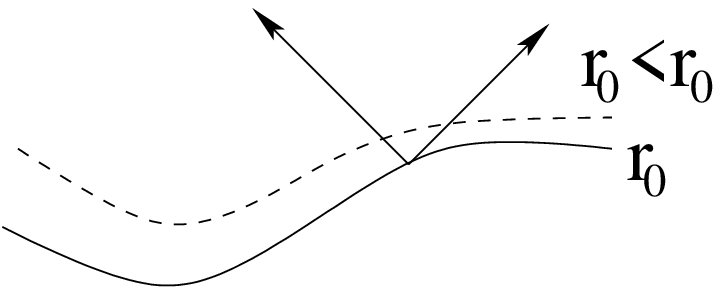}
  \caption{Constant area radius surfaces in spherical symmetry and
    null normal geodesic families, showing whether the surfaces are
    trapped or not.
    \label{RTrapped}}
\end{figure}

In general, it can be difficult to identify all trapped surfaces in a
given space-time, but spherical trapped surfaces in a spherically
symmetric spacetime are simple to detect. (See, e.g.,
\cite{TrappedVaidya,OuterTrappedVaidya} for non-spherical
trapped surfaces.) We use this here as an example to illustrate the
relation between the regions where trapped surfaces occur and
space-time singularities.  With a line element
\[
 \md s^2=-N(r,t)^2\md t^2+R(r,t)^2\md r^2+r^2\md\Omega^2
\]
formulated in the ``area radius'' $r$, we consider submanifolds
defined by $r=r_0$ being constant. This is usually a time-like
submanifold as illustrated in the left part of Fig.~\ref{RTrapped}.
While the ingoing null geodesic family moves toward smaller $r<r_0$
and is contracting, the outgoing null normal geodesic family moves to
larger $r>r_0$ and is expanding. Such a 2-sphere obtained as the cross
section of a time-like constant-$r$ surface is thus untrapped. When
the submanifold $r=r_0$ is space-like, by contrast, both the ingoing
and outgoing null normal geodesic families move to smaller $r$ (or
larger depending on whether the surfaces are future or past trapped);
see the right part of Fig.~\ref{RTrapped}. Any cross section of a
space-like constant-$r$ surface is thus trapped.

\begin{figure}
  \includegraphics[height=.1\textheight]{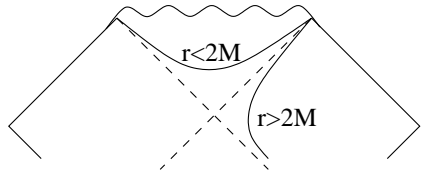}
  \caption{Conformal diagram of the Schwarzschild solution with
    time-like and space-like submanifolds of constant area radius.
    \label{Schwarzschild}}
\end{figure}

Thus, whenever $n_a=(\md r)_a$ is time-like, $g^{ab}n_an_b=R^{-2}<0$,
we have trapped surfaces. (If we use a line element with non-zero
shift $N_r=g_{rr}N^r$, the inequality reads
$g^{rr}=(R^2N^2-N_r^2)/N^2R^4<0$.)  For the Schwarzschild solution
$R^{-2}=1-2M/r$, and any sphere with $r_0<2M$ is trapped. To the
future of this region there must thus be a singularity, as it is drawn
in the usual conformal diagram Fig.~\ref{Schwarzschild}.

\subsection{General situation and Alternatives}

Singularity theorems demonstrate the stability under non-symmetric
perturbations of singularities explicitly seen in symmetric solutions,
such as the Schwarzschild or homogeneous cosmological solutions.
Historically, this played an important role in accepting the
importance of singularities for solutions of general relativity.  But
as in any mathematical theorem, assumptions are certainly necessary.
The most general theorems, those with the weakest assumptions, only
show that one incomplete geodesic exists. Moreover, the general
structure (such as curvature divergence) is not illuminated at all.
The significance of singularity theorems can thus be questioned, and
indeed non-singular (though maybe not fully realistic) solutions exist
even if positive energy conditions are assumed
\cite{NonSingSol,NonSingSolProof}.

Nevertheless, no general conditions for non-singular solutions are
known, and thus singularities cannot be ignored in general relativity.
Generic solutions have boundaries which cannot be penetrated by
geodesic observers provided that positive energy conditions hold.
General relativity is thus incomplete as it does not show what happens
at and beyond boundaries of its solutions.  Extensions of the theory
are necessary.  The key to solving the singularity problem is not to
find non-singular solutions but to provide sufficiently general
conditions under which non-singular behavior would be guaranteed. This
is not available in classical gravity even if no energy conditions are
imposed at all.

Rather than using geodesic completeness as a criterion for
non-singular behavior, one can use alternative conditions for
non-singular space-times.  One that is more physically motivated is
generalized hyperbolicity \cite{GenHyp} which states that all
space-times allowing a well-posed initial value problem for standard
matter fields are to be considered non-singular.  This uses physical,
potentially fundamental fields rather than test particles. It is more
general since (conical space-time) solutions are known which are
geodesically incomplete but satisfy generalized hyperbolicity
\cite{GenHypCon,GenHypHyp}. However, no general results are available
at present while the usual cosmological and black hole singularities
certainly present counter-examples.

Instead of changing criteria for singularities, modifying gravity
might lead to better situations: examples include alternative degrees
of freedom such as test strings rather than particles in string theory
or properties of quantized space-time in quantum gravity as they will
be discussed in the next section.

\section{Beyond General Relativity}

Singularity theorems demonstrate geodesic incompleteness under quite
general assumptions, based mainly on differential geometry (the
geodesic deviation and Raychaudhuri equations) in combination with
positive curvature. This implies focusing of geodesic families which,
together with topological conditions, result in space-time
singularities.  But special non-singular solutions do exist
classically, so the question is not if one can avoid the conclusions
by evading assumptions.  What is missing is a general mechanism by
which one can conclude non-singular behavior in a sufficiently general
class of physical situations.

Violating energy conditions is an obvious candidate to evade the usual
singularity theorems, but even this does not work generally enough for
the types of matter we seem to need for fundamental physics.
Singularity theorems have been proven, with different assumptions, for
instance in the context of inflaton fields which violate positive
energy conditions \cite{InflSing}. Moreover, dropping positive energy
assumptions does not necessarily improve the situation but usually
makes it worse as the development of sudden future singularities in
so-called phantom matter field models shows \cite{FutureSing}.

Other than that, only modifications of gravity itself rather than
matter can help. But also this will be subtle because not much of
general relativistic dynamics is being used in singularity theorems.

\subsection{Facets of the singularity problem}

The singularity problem is a complicated issue to be addressed in a
more general theory of gravity, extending general relativity in a
well-defined form. For any explicit discussion, the main difficulties
are:
\begin{enumerate}
\item No general classification of singularities is available and many
  different types exist.
\item Singularities are not always accompanied by unbounded curvature
  as in the best known examples.  It is thus not sufficient to address
  only unbounded curvature because this is not even shown in
  singularity theorems.
\item In fact, not all singularities should be resolved
  \cite{SingValue}. Some are useful to rule out negative mass, e.g.\
  of the Schwarzschild solution. Such singularities are typically
  time-like rather than space-like which is an important additional
  property not covered in singularity theorems.
\end{enumerate}
In isotropic models, one can often construct bounce pictures where the
volume of any solution is bounded away from zero. This can be achieved
by appropriate modifications of the dynamics which avoid unbounded
curvature. But such models do not serve as a general mechanism.

Curvature singularities, although not necessarily implied by the
singularity theorems, are the best known types realized in cosmology
and black hole physics. They can often be dealt with in special ways
which more or less directly ensure bounded curvature. In particular,
large curvature implies high energy regimes of field theories defined
on a singular gravitational background. As usually, field theories are
expected to receive strong quantum corrections in high energy regimes
which, when even the gravitational field is at ``high energies'' in
the sense of strong curvature, may also modify gravitational dynamics
itself. General relativity would then only be obtained as an effective
description, or the small curvature limit of a suitable extension
valid also at high curvature.

In quantum field theory and condensed matter physics, effective
descriptions on small energy scales or large length scales can be
defined by integrating out ``massive'' or short wave length degrees of
freedom which will become important at high energies. At low energies
those modes can safely be ignored, but they will become relevant when
typical energies reach their mass or when the curvature radius
approaches their length scale. Ignoring the additional degrees of
freedom in such a situation implies deviations, or possibly a failure
of the low energy description in the form of singularities.  It is
then only the effective description which fails and appears to be
singular while the more fundamental theory can (and should) be
non-singular.

This is an interesting and partially successful picture for which many
examples are available. Most of these examples are motivated by
special known solutions (mainly Friedmann--Robertson--Walker ones) and
devised with cosmological bounces in mind. So far, they are not
general enough to extend to more complicated, especially
inhomogeneous, solutions. Moreover, they do not address the true
problem of gravitational singularities: not all singularities have
large curvature, and diverging curvature is not the basic mechanism
behind singularity theorems. By addressing exclusively unbounded
curvature one appears to be treating a symptom rather than the cause.

\subsection{Example: string theory}

String theory \cite{String} provides an example for a theory whose
dynamics reduces to that of general relativity for small curvature and
low energies but differs at large energies. In particular, string
theory can be quantized perturbatively which is not the case for
general relativity without high energy corrections. Conceptual
features are, however, quite different in string theory. For instance,
while solutions of general relativity are not just a metric
perturbation but rather the space-time manifold itself, as discussed
before, string theory in its present version formulates gravitational
excitations on a given metric background.  This certainly has
implications for how generally the singularity issue can be addressed,
keeping in mind that singularities in general relativity are
understood as boundaries of space-time arising through the dynamical
laws that govern its own structure. If a background manifold is put in
from the outset, a discussion and resolution of singularities at a
general level becomes impossible.  String theory does, however,
provide valuable insights into how special singularities can be
resolved by new degrees of freedom. This occurs mainly through a
different viewpoint on test objects replacing geodesics, and through
candidates for massive degrees of freedom not contained in general
relativity. Many examples are discussed, e.g., in the recent review
\cite{Craps} to which we refer for details and more complete
references.

In string theory, one uses 1-dimensional strings or higher dimensional
branes rather than pointlike particles as basic objects. World-volumes
of test strings or branes then replace geodesics followed by point
particles as the submanifolds whose incompleteness would signal a
singularity in the spirit of general relativity's singularity
theorems. This has been shown to change completeness results and lead,
in this sense, to more regular behavior especially for conical or
orbifold singularities (which are not space-like). It also provides
examples for new degrees of freedom which, when taken into account
rather than being integrated out, are necessary for regular behavior
\cite{Conifold}.

This, however, has been difficult to extend to dynamical space-times
such as those displaying curvature singularities. General arguments
why such space-times provide a qualitatively different challenge have
been presented in \cite{StringSing}, and several so far unsuccessful
alternative attempts can be found in
\cite{NullOrbifoldI,NullOrbifoldII}. The main difficulty is that
string perturbation theory generally breaks down in such space-times.
(Perturbation theory might be useful for the singularity issue in the
context of tachyon condensation \cite{Tachyon}, but has so far been
used only for null singularities where curvature does not diverge.)

A second source of additional degrees of freedom which could become
light close to a classical singularity and help to resolve it are
string winding modes around topologically non-trivial components of
space-time (or brane separation parameters): their mass $m\propto
R^{-1}$ is proportional to the inverse radius (measured in the
background metric the strings are propagating in) of the compact
direction they are winding around. This is highly massive, and thus
negligible in low energy effective theory, if extra dimensions are
small, but can become relevant in high curvature regimes.  It has been
shown that winding modes can easily lead to bounces \cite{StringGas},
but typically only of some directions. With small extra dimensions one
is necessarily dealing with anisotropic geometries such that a bounce
in one direction does not imply a spatial volume bounded away from
zero. Usually, only the compact dimension bounces. Moreover, by design
special topologies or configurations are required for winding modes to
exist which spoils prospects for a general mechanism as gravitational
singularities also occur in simply connected space-times.

These constructions referred to strings propagating in non-evolving
backgrounds. It is much more difficult to find analogous mechanisms in
dynamical space-times as they arise in cosmology or in the interior of
black holes close to their singularities. In such a context, not only
technical difficulties arise but also applications of low-energy
effective actions, which are mainly being used to study the effect of
massive degrees of freedom, are not always general enough
\cite{Karpacz}. While low energy effective actions are well-suited to
study the propagation and scattering of fields which are not highly
excited out of their vacua, dynamical space-times in quantum gravity
have to include a gravitational state far away from its vacuum. Then,
more general effective equations are necessary which, requiring good
knowledge of the quantum gravity state, are more complicated to
derive.

The main difficulty is that in all examples one is still using test
objects in a background space-time, such as new fields provided by
positions of branes. This does not include the dynamics of space-time
itself. Moreover, strong back-reaction effects occur
\cite{SingBackReact,NullOrbifoldII} showing that gravitational
dynamics is very relevant and that a pure background treatment is
insufficient. The most detailed scenario in this context has been
formulated for the Schwarzschild-AdS singularity, rather than a
dynamical cosmological one. From an analysis of correlation functions
of the conformal boundary field theory one can conclude that bulk
properties such as horizons and singularities only emerge in the
classical limit but are not present in quantum gravity
\cite{SchwarzschildAdSNonSing}. 

Tight arguments have been put forward which indicate that bounces
found in homogeneous models are unlikely to extend to inhomogeneous
situations \cite{StringInHom}. Although they have been discussed there
mainly in the context of the AdS/CFT correspondence, some of the
arguments are general enough to caution against direct generalizations
of homogeneous results in any context, not just in string theory.
Specifically, an upside-down potential for field modes is argued to
arise in a boundary field theory description of the cosmological
situation. Field modes in this potential will reach infinity in a
finite amount of time, corresponding to the classical singularity.
Since all modes are independent, they behave differently even if they
started out in a highly correlated manner from an isotropic initial
configuration. Quantum gravity, by way of a self-adjoint extension of
the field Hamiltonian which leads to reflecting boundary conditions at
infinite values of the fields, could make the behavior non-singular.
But if this happens, the field modes are unlikely to return to an
highly correlated state as the initial one. Thus, while evolution of
the quantum theory continues, it does not easily lead to a classical
bounce back to a classical geometry.  The main property of
inhomogeneities used in this argument is the large number of
fundamental degrees of freedom, which all need to be re-excited
collectively in a special way for a smooth bouncing geometry to
result. Independently of the specific quantization of inhomogeneities,
this is much easier to achieve in homogeneous models with a small
number of degrees of freedom than in inhomogeneous ones.

\subsection{Geometry}

Singularity theorems are mainly statements about differential geometry
as they refer to properties of geodesics on a curved manifold.
Einstein's equation is used only at one place, relating positive
energy to positive curvature, which then implies focusing effects in
the Raychaudhuri equation. Focusing through positive energy is used in
the most common theorems, but is not the only reason since
singularities easily arise with violated energy conditions.

For a general solution of the singularity problem one should thus
focus on geometry, not just on dynamics. Although both are intertwined
in general relativity, geometry determines which type of dynamics is
possible such that it can be viewed as more basic. Rather than
modifying gravitational dynamics in a given geometrical setting, using
a different geometry could be much more successful to avoid the
far-reaching conclusions of singularity theorems.

An alternative geometry is automatically provided by background
independent quantizations, such as canonical quantizations. Such
theories are not based on objects in a background space-time but they
quantize full metric components as the non-perturbative dynamical
objects. For instance, most versions employ wave functions supported,
e.g., on the space of spatial metrics $q_{ab}$ which is the
configuration space of canonical general relativity. Geometrical
objects then become operators acting on these wave functions with
properties generally very different from classical smooth geometry.  A
new quantum geometry underlying gravity then arises with an entirely
new setting for the singularity issue. Despite of the difference to
classical geometry, for any consistent quantum theory of gravity
smooth space-times have to be approached as a classical limit far away
from singularities. But large deviations from classical behavior can
occur around classical singularities, possibly resulting in regular
equations.  Quantum geometry then, if this picture is successful,
provides links between classical parts of space-time which would
otherwise be interrupted by singularities.

Canonical techniques treat space and time differently because time
derivatives of fields are replaced by momenta but their spatial
derivatives are retained. This is also true in relativistic theories
where manifestly covariant theories are rewritten canonically in a
form referring separately to space and time. Although this hides the
covariance of field equations, constraints ensure that solutions still
respect the equivalence principle.

\begin{figure}
  \includegraphics[height=.15\textheight]{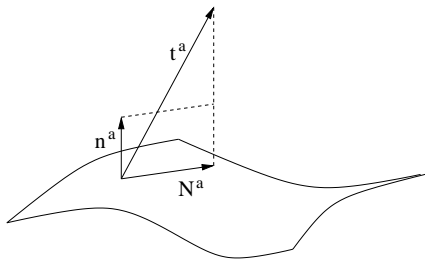}
  \caption{Decomposition of the time evolution vector field $t^a$ into
    parts normal and tangential to a spatial manifold.
    \label{Split}}
\end{figure}

In general relativity, a canonical formulation is defined by
introducing a foliation of space-time into a family of spatial slices
in terms of a time function $t$ such that the slices are
$\Sigma_t\colon t={\rm const}$. Moreover one chooses a time evolution
vector field $t^a$ such that $t^a\nabla_at=1$ which determines how
points on different spatial slices are identified (along integral
curves of $t^a$) to result in spatial fields ``evolving'' in
coordinate time $t$. In addition to the time evolution vector field
there is the geometrically defined unit normal vector field $n^a$ to
the spatial slices in space-time. This allows one to decompose the
time evolution vector field $t^a=Nn^a+N^a$, as illustrated in
Fig.~\ref{Split}, into a normal part, whose length is given by the
lapse function $N$, and a tangential part given by the spatial shift
vector $N^a$.  One can then split off the time components from the
inverse metric $g^{ab}$ by defining the inverse spatial metric
$q^{ab}=g^{ab}+n^an^b$ such that $q^{ab}n_b=0$. Thus, the spatial
metric indeed has only components tangential to spatial slices.
Solving the relation for
\[
 g^{ab}=q^{ab}-n^an^b=q^{ab}-\frac{1}{N^2}(t^a-N^a)(t^b-N^b)
\]
and inverting it results in the space-time metric (\ref{CanonMetric})
in the canonical form already used before. As is clear from the
construction, lapse function and shift vector arise through the
space-time foliation and define coordinate time evolution through the
way different spatial slices are identified. They are thus related to
the space-time gauge but are not dynamical fields. This confirms the
realization that they are not subject to evolution equations as their
time derivatives do not occur in Einstein's equation.

Evolution equations do result for the spatial metric $q_{ab}(t)$,
interpreted as a time dependent field through its values on different
$\Sigma_t$. In a canonical formulation we have first order equations
(\ref{qdot}) and (\ref{pdot}) for $q_{ab}$ and its momentum, related
to extrinsic curvature $K_{ab}=\frac{1}{2N}( {\cal
  L}_tq_{ab}-2D_{(a}N_{b)})$ of $\Sigma_t$ (with ${\cal L}_t$ denoting
the Lie derivative along $t^a$).  These phase space coordinates given
by spatial metric components $q_{ab}$ and their momenta are subject to
constraint equations (\ref{constr}) implementing their
dynamics, as we will discuss in more detail in the next section. For now,
it suffices to know that, when quantized, wave functions on the space
of metrics (if a metric representation is chosen) arise which are
subject to differential or difference (diff.\footnote{Whenever we
  write ``diff.'' we mean that differential or difference equations
  can occur depending on the context (the quantization scheme used)
  but the statement does not depend on the type of equation.})
equations depending on the quantization scheme.

There are then no test objects in a background space-time, but wave
functions or gravitational observables are the fundamental dynamical
entities. Differential geometry becomes applicable only in a classical
approximation to describe space-time, and only in classical regions
would geodesics be defined at all.  Geodesic incompleteness is thus
inapplicable as a criterion for singularities in quantum gravity. This
suggests a natural answer to why incompleteness occurs so generally in
general relativity: geodesics themselves are only valid as long as a
differentiable classical geometry can be assumed. (In fact, there are
cases of geodesic incompleteness where simply the metric ceases to be
differentiable. These are typically the examples where generalized
hyperbolicity as a criterion leads to non-singular space-times
although they would be geodesically incomplete.) When quantum geometry
becomes relevant, geodesics stop and have to be replaced by something
more appropriate. Only classical geometry would end, but not quantum
gravity with its own version of quantum geometry. Such a scenario
looks promising but it has to be developed and verified in detail,
requiring explicit candidates for background independent quantum
gravity such as a canonical quantization. The question then remains:
How do we address or even define singularities in such a context?

\subsection{Quantum hyperbolicity}

Classical singularities refer to properties of spacetime as a metric
manifold. Abstractly, they are thus identified through properties of
the metric tensor on certain submanifolds, although an explicit
classification in general terms is lacking.  In a canonical
quantization, classical singularities must then correspond to
properties of states, or of suitable observables in a Heisenberg
picture. The usual example is how a wave function in the metric
representation is supported on certain submanifolds of the space of
spatial metrics which classically would imply a singularity. Before we
can even use quantum dynamics we must be able to provide an
unambiguous one-to-one correspondence between classical singularities
and metric tensors. This could to some extent be done in terms of
curvature invariants, but must be more general since not any
singularity refers to curvature. Such a classification is not
available in full generality and is thus the major difficulty in any
general discussion of the singularity problem in quantum gravity,
irrespective of the specific quantization approach followed.

\begin{figure}
  \includegraphics[height=.15\textheight]{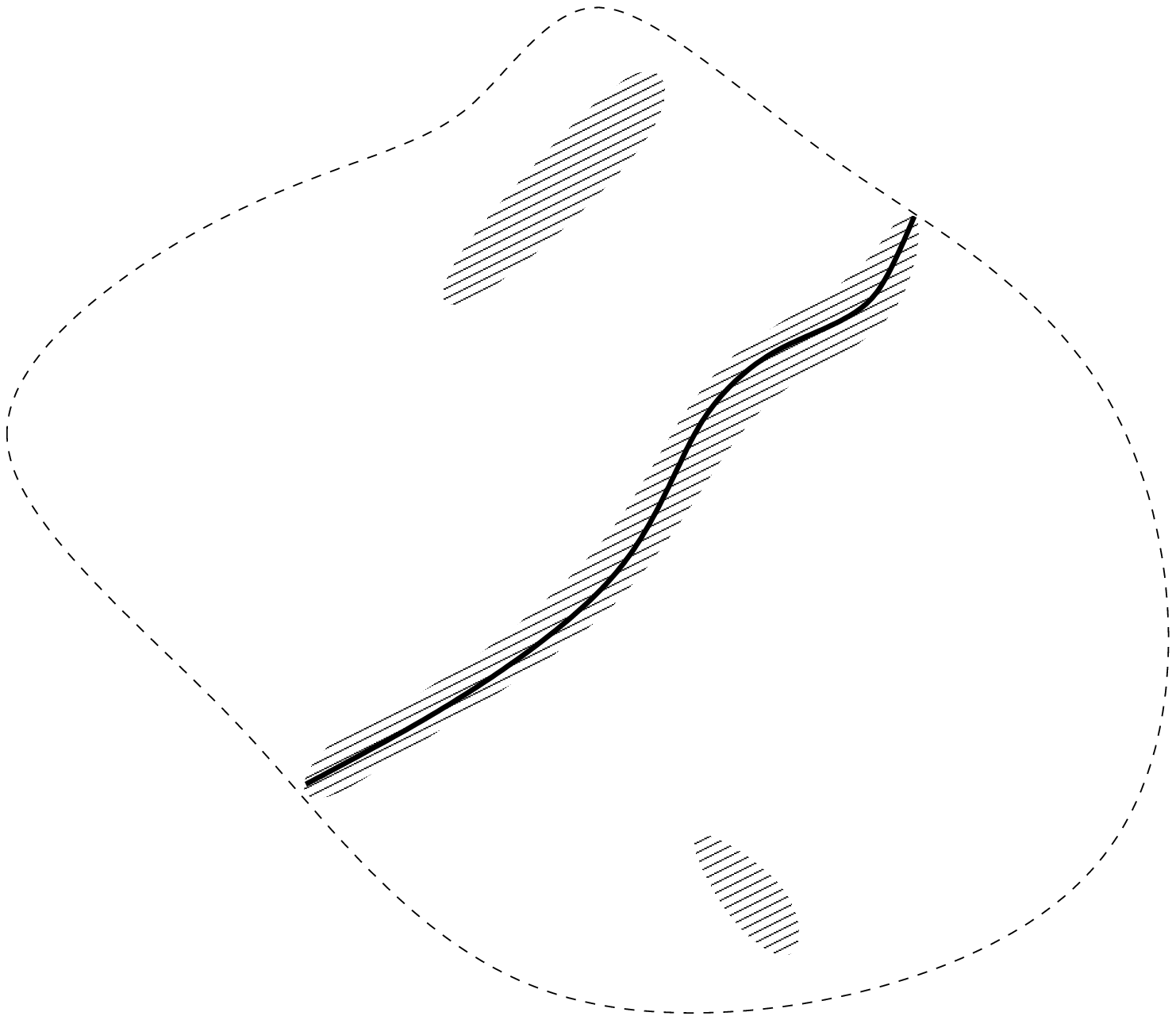}
  \caption{A sketch of some part of superspace with classically singular
    regions. They could be finite regions or entire submanifolds
    splitting superspace into several disconnected components. The
    latter are most dangerous because they prevent physical
    information to be extended uniquely across that boundary. In
    addition, singular regions may be sharply defined submanifolds or
    washed-out regions due to quantum uncertainty. A detailed
    classification is not available in full general relativity but can
    often be completed in models which assume symmetries.
    \label{SuperSing}}
\end{figure}

Although a general classification is not available, a new criterion
for gravitational singularities becomes possible: If a state can
uniquely be extended across or around all submanifolds of classically
singular configurations they do not pose obstructions to quantum
evolution in any sense. If, however, a state such as a wave function
on the space of metrics cannot be extended uniquely across a
classically singular submanifold, there would still be a boundary to
quantum evolution and quantum space-time would remain incomplete. This
viewpoint is the natural extension of generalized hyperbolicity from
matter fields on space-time to the fundamental object of quantum
space-time itself. We therefore call it the {\em principle of quantum
  hyperbolicity}.

A verification in concrete ways requires quantum geometry for the
location of potential singularities and quantum dynamics to see if
states can uniquely be extended across such places.  The set-up in this
form is general without specific reference to unbounded curvature. It
can potentially deal with the general types of singularities
implied by the theorems.

States, as solutions to the dynamical equations of quantum gravity,
thus have to be extended uniquely across classical singularities.  As
a general principle, it entails several sub-issues necessary for its
verification:
\begin{itemize}
\item First, phase space locations of classical singularities in
  terms of metric or curvature components have to be unambiguously
  identified.
\item The structure of submanifolds in phase space corresponding to
  classical singularities according to the first point then determines
  how the classical space of metrics is separated into several
  disconnected components. In some sense to be specified, any state
  has to be extended uniquely between any two such regions. This gives
  meaning to an extension of states across classical singularities.
\item Before this extension can be shown to exist uniquely,
  representations of states or relational observables must be chosen.
  The formulation of an extension is necessarily representation
  dependent because, for instance, no background coordinates are
  available in which classical metrics $q_{ab}(t)$ would approach a
  singularity.  Any extension is thus provided in phase space
  variables (internal time) which are monotonic around a classical
  singularity. This selects representations in which one observes an
  extension of a wave function from one part of its support to
  another, rather than a superposition of different branches.
\item Finally, the dynamical equations of quantum gravity will be used
  to verify that all states can be extended in the way envisioned in
  earlier steps. For a general answer to the singularity issue it is
  important to prove this extension for all allowed states rather than
  for restricted classes such as those given by one specific initial
  condition.
\end{itemize}

For a specific proposal of how quantum hyperbolicity could be
realized, one can use the usual Wheeler--DeWitt type wave functions on
the space of metrics \cite{QCReview,DeWitt}. Then, metric variables
would be used to provide extensions across a classical singularity.
The first point above, classifying classical singularities through the
metric behavior, cannot be performed currently in full generality, but
is often easily available in symmetric models. One thus studies mini-
or midisuperspaces of metrics respecting a certain symmetry and
possible extensions of wave functions within this class of metrics.
Also quantum dynamics can often be obtained explicitly in such models
such that the program can be carried through. Also here, one has to be
careful with interpreting results in any given model because the
structure of classical singularities as well as quantum dynamics are
truncated by restricting oneself to one class of metrics. But the
investigation of several different models, covering different types of
classical singularities, provides valuable information as to whether
or not a general mechanism providing quantum hyperbolicity can exist.
It also shows how details of the specific quantum theory of gravity
used in the process matter which allows conclusions about general
constructions.

If realized, quantum hyperbolicity provides a quantum version of
generalized hyperbolicity and thus deals with the well-posedness of
evolution problems. If quantum equations of wave functions are
well-posed, there are no boundaries for quantum evolution even where
one would classically expect a singularity. Given that issues of
hyperbolicity are very difficult in general relativity and generalized
hyperbolicity at the classical level has not been studied much one
could question the feasibility of quantum hyperbolicity as a
verifiable criterion in quantum gravity. After all, quantum dynamics
is expected to be much more complicated than classical dynamics from a
mathematical point of view. In fact, quantum dynamics in background
independent approaches has not even been fully formulated yet.
Nevertheless, quantum theory allows an important simplification of
testing well-posedness: If one uses a Schr\"odinger picture of states,
dynamical equations are {\em linear}. This removes one of the major
difficulties present in classical general relativity. The complicated
issue is, of course, to extract observable information out of the
states where the non-linearity of gravity enters. Moreover, solutions
even to linear quantum equations can be difficult to find explicitly
in inhomogeneous models with many independent degrees of freedom.  But
for hyperbolicity we only need to study the well-posedness of initial
or boundary value problems without the need of having explicit
solutions available. This is much more feasible for linear compared to
non-linear equations. Then, given that states can be uniquely extended
across classical singularities once quantum hyperbolicity has been
verified, one is assured that observable information extracted from
such extended states also reaches from one disconnected part of
classical superspace to others.  This is a hopeful sign that general
statements about the singularity issue can be made in quantum gravity
without having to face all the difficult issues, just as singularity
theorems were derived in classical gravity without much knowledge
about the general solution space.

\subsubsection{Examples: isotropic, homogeneous space-times and the
  BKL conjecture}

Nevertheless, simple examples demonstrate that there are many
non-trivial issues in verifying quantum hyperbolicity.
The general idea is best illustrated in isotropic models with a single
metric component given by the scale factor $a>0$, subject to the
Friedmann equation $(\dot{a}/a)^2=\frac{8\pi G}{3}
\rho(a,\varphi)$. Classical singularities for the usual matter
contributions, such as a scalar field,
are reached at $a=0$. This gives a simple and general identification
of classical singularities for isotropic geometries. Since $a$ is
positive, the classically singular submanifold is a boundary rather
than an interior submanifold. Wheeler--DeWitt wave functions in the
metric representation take the form $\psi(a,\varphi)$ with a scalar field
$\varphi$ assumed as the matter content.

This provides the simplest situation, but is very special.  Even
keeping homogeneity, the situation changes considerably if anisotropy
is allowed such as in the Bianchi I model. There are then three
independent metric components $a_I>0$, $I=1,2,3$, determining a metric
$\md s^2=-\md t^2+\sum_Ia_I^2(\md x^I)^2$. Instead of the Friedmann
equation, we have the constraint
$\dot{a}_1\dot{a}_2+\dot{a}_1\dot{a}_3+ \dot{a}_2\dot{a}_3=0$ which is
solved by the Kasner solutions $a_I\propto t^{\alpha_I}$ where
$\alpha_I$ are real numbers such that
$\sum_I\alpha_I=1=\sum_I\alpha_I^2$. These two equations have only
solutions (except for the non-dynamical one corresponding to Minkowski
space) satisfying $-1<\alpha_I\leq 1$. One of them must then be
negative while the other two are positive. The total volume is
proportional to $a_1a_2a_3\propto t$ which vanishes at $t=0$,
corresponding to the classical singularity. Again, we have a simple
characterization, but now the behavior of the metric components is
very different: one metric component diverges at the singularity. The
singular submanifold is thus not only a boundary but located at
infinity. This makes investigating the well-posedness of initial
value problems of wave function $\psi(a_1,a_2,a_3)$ in a neighborhood
of the classical singularity more complicated. 

In any case, it demonstrates that conclusions drawn from isotropic
models cannot easily be generalized because the structure of classical
singularities themselves changes.  On the other hand, there are
indications related to the BKL conjecture \cite{BKL} that anisotropic
models are quite generic and provide crucial information even for
inhomogeneous situations. In this context one is looking for generic
asymptotic solutions close to a space-like curvature singularity at
$\tau=0$, which are argued to be of the form
\[
\md s^2 = -\md\tau^2+\tau^{2\alpha_1(x)}(\omega_1)^2+
\tau^{2\alpha_2(x)}(\omega_2)^2+ \tau^{2\alpha_3(x)}(\omega_3)^2\,.
\]
Thus, a homogeneous Bianchi model, whose invariant 1-forms
$\omega_a^I$ are used in the spatial part of the metric, is
generalized by allowing position dependent Kasner exponents
$\alpha_I(x)$. For a Bianchi I model, for instance, one would simply
have $\omega^I=\md x^I$ in terms of Cartesian coordinates $x^I$. More
generally, the 1-forms $\omega^I_a$ are left invariant 1-forms on a
3-dimensional Lie group and thus satisfy the Maurer--Cartan relations
$\md\omega^I=-\frac{1}{2}C^I_{JK} \omega^J\wedge\omega^K$ with the
structure constants $C^I_{JK}$ of the Lie algebra. 

Since inhomogeneities in this class of space-times are modeled by
space dependent exponents $\alpha_I(x)$, there are gradient terms
containing the spatial derivatives $\partial_a\alpha_I$ added to the
homogeneous equations of motion. This can be derived from the
Christoffel connection which, in addition to the terms occurring in a
homogeneous model with constant $\alpha_I$, receives contributions
\[
 \delta{}^{(3)}\Gamma^c_{ab}= \frac{1}{2} q^{cd}(\partial_a \delta
 q_{bd}+ \partial_b \delta q_{ad}- \partial_d \delta q_{ab})
\]
where $\delta q_{ab}\md x^a\md x^b= \sum_I
(\tau^{2\alpha_I(x)}-\tau^{2\alpha_I}) (\omega^I)^2$ is the difference
spatial metric after introducing inhomogeneity. This term can easily
be computed:
\begin{equation} \label{Grad}
 \delta{}^{(3)}\Gamma^c_{ab}= \log\tau
 (\delta^c_b\partial_a\alpha_{(b)}+ \delta^c_a\partial_b\alpha_{(a)}-
 \delta^{cd}\delta_{ab}\partial_d\alpha_{(a)})
\end{equation}
which diverges logarithmically at $\tau=0$. Thus, 3-curvature terms
of $^{(3)}R_{abc}{}^d$ resulting from this contribution, diverge as
$(\log\tau)^2$, and the 3-Ricci scalar as
$\tau^{2\alpha_I}(\log\tau)^2$.

This is to be compared to contributions of the homogeneous curvature
scalar to see if inhomogeneities play a role for curvature and
equations of motion. For a general Bianchi class A model with structure
constants parameterized in the form
$C^I_{JK}=\epsilon_{NJK}n^{(I)}\delta^{IN}$ \cite{MacCallum} we have
the 3-Ricci scalar
\begin{equation} \label{R3}
 {}^{(3)}R_{\rm hom}= -\frac{1}{2}\left(\frac{n^1q_1}{q_2q_3}+
   \frac{n^2q_2}{q_1q_3}+ \frac{n^3q_3}{q_1q_3}- 2\frac{n^1n^2}{q_3}-
   2\frac{n^1n^3}{q_2}- 2\frac{n^2n^3}{q_1}\right)
\end{equation}
for a diagonal metric of the form $q_{IJ}=q_{(I)}\delta_{IJ}$ in the
$\omega^I$-basis. Assuming, without loss of generality, that
$\alpha_1<0$ is the negative one of the Kasner exponents, the
strongest divergence of ${}^{(3)}R_{\rm hom}$ is given by
$q_1/q_2q_3\sim \tau^{2\alpha_1-2\alpha_2-2\alpha_3}$. Since
$2\alpha_1-2\alpha_2-2\alpha_3<2\alpha_I$ for all $I$ if $\alpha_1$ is
the negative exponent, the homogeneous contribution (\ref{R3}) is more
divergent than the gradient contributions resulting from (\ref{Grad}).
This gives rise to the conjecture that spatial derivatives are
subdominant asymptotically close to the singularity and that a general
solution behaves locally as a homogeneous model of the Bianchi IX form
($n^1=n^2=n^3=1$) where all diverging terms in ${}^{(3)}R_{\rm hom}$
are present.  

However, the BKL conjecture, although by now strongly supported
\cite{Garfinkle,WS:AR}, has not been proven, and it is even less
secure that it will extend to a quantum version. Diverging curvature
is relevant in the argument, and it thus refers explicitly to
curvature singularities. Classical equations of motion have been used
which themselves may change in such strong quantum regimes. Moreover,
one must get close to the classical singularity in order to have
subdominant spatial gradients. Bounce models, as they are often
discussed, avoid this high curvature regime and cannot appeal to the
BKL conjecture to justify a possible extension to inhomogeneous
situations.  Thus, inhomogeneous models as close to the full situation
as possible must still be considered in detail.

\subsubsection{Boundaries}

As seen in the examples, when metric variables are used for a
characterization, singularities usually appear at boundaries of
superspace since $\det q_{ab}>0$ by definition and singularities often
have degenerate metric or triad components. This makes it difficult to
extend states across such submanifolds; one could at most ask that the
state at the classical singularity be well-defined and that, in a
certain sense, the boundary does not pose an obstruction to physical
evolution. This suggests, as it is often used, to impose boundary
conditions on wave functions right at a classical singularity, with
different motivations for specific proposals
\cite{DeWitt,nobound,tunneling,SIC}.  If these are conditions which
imply, for instance, wave packets being reflected off the boundary,
one can argue that evolution is not interrupted. Such an
interpretation is, however, not easy to justify in general. It is
dependent on choices of the precise form of boundary conditions and
difficult to make generic beyond isotropic models.  Even if the wave
function is not supported at a classical singularity, it does not mean
that all extracted physical information is regular. To test this one
would have to construct observables and the physical inner product in
order to compute expectation values. If all these quantities remain
regular, one can conclude that the classical singularity has been
resolved. But if only wave functions have been shown, or restricted by
boundary conditions, to remain regular one is not in a position to
decide about singularity resolution. For instance, while a wave
function constrained to vanish at $a=0$ of an isotropic model taken
together with the usual probability interpretation (which is itself
subject to difficult interpretational issue in quantum gravity)
implies directly that any physical quantity is supported only away
from zero, it does not lead to general lower bounds for $a$. While in
any given state the expectation value of $\hat{a}$ in the usual inner
product would have a positive lower bound, one can always choose a
state where the expectation value comes arbitrarily close to zero. The
boundedness of $a$ is then only put in through selecting the
particular boundary condition, but cannot be regarded as a consequence
of quantum gravity.  Moreover, the physical inner product in which one
computes the expectation value might itself contribute a function of
$a$ diverging at $a=0$. While one can sometimes compute the physical
inner product in specific models, which does not only involve assuming
symmetries but also selecting the matter content, there is no general
information on its behavior. Any interpretation of properties of wave
functions which relies on the physical inner product is thus highly
difficult to turn into a general argument to address the singularity
issue.

Probably the most advanced discussion in this context can be found in
the application \cite{UnitaryCollapse,QuantumCollapse} to
gravitational collapse. It is shown there that a canonical
quantization of the explicitly determined reduced and deparameterized
phase space (i.e.\ an internal time has been introduced to describe
physical evolution), which solves all the constraints, is non-singular
in the sense that an initial wave packet not supported at the
classical singularity always vanishes at the classical singularity.
This provides a full analysis making use of Dirac observables and a
physical inner product. It is by far the strongest result in this
context. Nevertheless, the question of whether it can be extended to a
general mechanism remains open.  First, the direct use of Dirac
observables, although necessary to discuss singularity removal
possibly implied by boundary conditions of wave functions, makes it
difficult to see what happens in other situations where Dirac
observables are rarely known explicitly.  The very fact that
Dirac observables can be computed explicitly in this model may imply
that it is quite special, including its dynamics.  But more
importantly, the mechanism in the collapse model depends on
assumptions on a semiclassical initial state: One must assume that the
initial state is not supported at all at the classical singularity but
only at a peak value far away. Then, unitary evolution preserves the
boundary condition which guarantees that the wave function will never
be supported at the classical singularity. While this looks innocent,
it makes singularity removal in this scheme unstable: If there is only
a tiny contribution to the wave function not vanishing at the
classical singularity, it would not change much of the initial
semiclassical behavior. But this already spoils the preservation of
the boundary condition and in general a wave packet approaching the
classical singularity will not stay away from it completely. The
mechanism, like others based on boundary conditions, is thus sensitive
to precise details of physics at the Planck scale which have to be
dealt with by extra assumptions.

Classical singularities located at boundaries of the space of metrics
thus do not offer an obvious general and verifiable way for regular
behavior. Quantum hyperbolicity does not require explicit observables
but can be formulated directly for general states solving the
dynamical constraints. Nevertheless, a location of classical
singularities at boundaries seems to prevent the applicability of
quantum hyperbolicity in any realistic sense because the choice of
boundary data matters crucially. But one should note that the
characterization of classical singularities as kinematical boundaries
in this way depends on the variables used. The main reason for the
location at boundaries was the restriction $\det q_{ab}>0$ which
obviously has to be imposed on the spatial metric. But geometry can
just as well be described in triad variables, which is even necessary
if fermionic fields are present.  If a co-triad $e_a^i$ is being used,
defined as three co-vector fields $e_a^i$, $i=1,2,3$ such that $\sum_i
e_a^ie_b^i=q_{ab}$, the sign of $\det e_a^i$ is relevant since it
determines the orientation of space while $\det q_{ab}$ will still be
non-negative. A surface $\det e_a^i=0$ is then interior, not a
boundary.

A quick look at the models studied before reveals that this sometimes
helps in rendering classical singularities interior submanifolds. In
isotropy, there is again a single co-triad component $e\in{\mathbb
  R}$, $e_a^i=e\delta_a^i$, such that $a=|e|$.  The classical
singularity at $e=0$ is then an interior point.  In these variables,
classical singularities are boundaries not simply by definition of the
basic variables but only for classical evolution which breaks down at
$e=0$. For quantum dynamics, the extendability of states thus becomes
testable.

In the Bianchi I model, however, we have $e_I\in{\mathbb R}$ such that
$a_I=|e_I|$. While the two vanishing triad components of a Kasner
solution would correspond to interior points, the diverging one
implies that the classical singularity is still located at the
infinite boundary of minisuperspace.  A further reformulation
alleviates this issue: we use a {\em densitized triad}
$E^a_i=|\det(e_b^j)|e^a_i$, with the triad $e^a_i$ inverse to $e_a^i$,
$e^a_ie_a^j=\delta_i^j$.  Then, the single densitized triad component
$p\in{\mathbb R}$ satisfies $a^2=|p|$ in isotropic models where the
classical singularity $p=0$ is still located in the interior. In
anisotropic models we have three components $p^I\in{\mathbb R}$
related to the metric components by $a_1=\sqrt{|p^2p^3/p^1|}$ and
cyclic. (There is a gauge transformation which changes the signs of
any two components $p^I$ leaving the third fixed. Thus, only
$\sgn(p^1p^2p^3)=\sgn\det E^a_i$ is physically distinguishable.)  For
a Kasner solution, $p^I\propto t^{1-\alpha_I}$ where $1-\alpha_I\geq
0$. All densitized triad components approach zero at an anisotropic
classical singularity which is thus realized as an interior point
\cite{HomCosmo}.

Typical homogeneous singularities occur as interior points of the
densitized triad space. This is also true for the Schwarzschild
singularity \cite{SphSymmSing} and employing the BKL conjecture one
can assume that general inhomogeneous singularities relevant for
cosmology and black holes may have the same behavior.  Given a candidate
for quantum evolution, the extendability of wave functions can then be
studied in finite neighborhoods.  The choice of variables thus matters
for quantum hyperbolicity which may seem an unwelcome dependence on
coordinatization. But it is not only the characterization of classical
singularities where the choice of variables matters but also, and even
more so, for the success of a chosen quantization scheme. This
determines which kind of quantum dynamics can be used with a set of
states represented on the space of metrics or triads. It is thus an
appealing possibility that the formulation of quantum dynamics may put
restrictions on the choice of basic variables in a form which may or
may not allow one to realize quantum hyperbolicity.  The structure of
classical singularities then becomes an important means to test
general issues of dynamics in quantum gravity in a way nicely
intertwining classical relativity with quantum geometry and dynamics.
We will discuss these issues in detail in the following sections.

\subsubsection{Requirements on quantum dynamics}

The quantum hyperbolicity principle is testable as it requires crucial
properties of quantum dynamics to be realized.  Continuations of wave
functions are relevant and thus the mathematical type of dynamical
equations. On the other hand, the principle is insensitive to
conceptual issues such as the interpretation of wave functions,
observables or evolution which are largely unresolved in quantum
gravity. Progress can thus be made on the singularity issue even
before quantum gravity is fully understood.

\begin{figure}
  \includegraphics[height=.1\textheight]{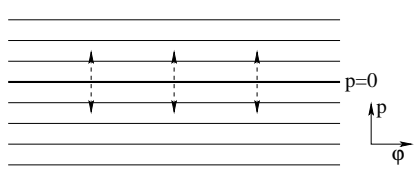}
  \caption{Isotropic minisuperspace in densitized triad variables $p$ with
    a matter field $\varphi$. Quantum hyperbolicity requires a unique
    extension of wave functions across the line $p=0$.
    \label{IsoCross}}
\end{figure}

Using densitized triad variables in isotropic cosmology leads to a
dynamical equation of the type $\hat{\Delta}\psi_p(\varphi)=
\hat{H}\psi_p(\varphi)$ with a diff.\footnote{Recall that this means
  differential or difference.} operator $\hat{\Delta}$ on superspace
and a matter Hamiltonian $\hat{H}$ which is usually a differential
operator on the matter field also containing metric components. For
quantum hyperbolicity we must be able, starting from suitable initial
values at one side of $p=0$ such as large positive $p$, to extend any
solution uniquely across $p=0$ as illustrated in Fig.~\ref{IsoCross}.

Any matter Hamiltonian $H$ defined with a fundamental field rather
than phenomenologically through an equation of state contains
$p^{-1}$, such as in
\[
 H=\frac{1}{2}|p|^{-3/2}p_{\varphi}^2+|p|^{3/2}V(\varphi)
\]
for a homogeneous scalar with momentum $p_{\varphi}$. The reason for
this general behavior is that momenta in a relativistic canonical
formulation carry a density weight. Since momenta appear in quadratic
form in usual matter Hamiltonians, this requires an inverse of the
determinant of the metric to get a well-defined, coordinate
independent integration of the total Hamiltonian. This inverse
determinant leads to $|p|^{-3/2}$ in an isotropic reduction.  For a
well-defined evolution interpretation of the equation in $p$, the
coefficients in the diff.\ equation must not diverge especially at
$p=0$. This requires that a quantization taking into account gravity
must lead to a matter Hamiltonian where the divergence of the kinetic
term at $p=0$ does not arise. One can interpret this as saying that
quantum hyperbolicity in an {\em isotropic model} requires bounded
curvature because classical curvature in an isotropic model behaves as
an inverse power of the scale factor. This requirement is intuitively
reasonable, but it is only secondary and derived from quantum
hyperbolicity for isotropic models.

This bounded isotropic curvature condition is much easier to check
than quantum hyperbolicity and thus provides an easily accessible
indication of its realizability. For the Wheeler--DeWitt equation (in
an ordering as it occurs, for instance, in
\cite{SemiClass,LoopCosRev})
\begin{equation} \label{WdW} \frac{2}{9}\ell_{\rm
    P}^4\frac{\partial^2}{\partial
    p^2}(\sqrt{|p|}\psi(p,\varphi))=\frac{8\pi
    G}{3}\left(\frac{\hbar^2}{2} |p|^{-3/2}
 \partial^2\psi/\partial\varphi^2- |p|^{3/2}V(\varphi)\psi(p,\varphi)\right)
\end{equation}
with the ordinary quantization of $p$ as a multiplication operator the
condition is obviously not satisfied: the right hand side diverges
generically at $p=0$. Then, even with the classical singularity being
interior in densitized triad variables, there is a break-down of the
initial value problem at $p=0$. Wheeler--DeWitt quantum cosmology
violates quantum hyperbolicity.

Discrete approaches provide simple solutions to this problem if the
continuous $p$-space is replaced by a lattice not containing $p=0$.
Coefficients of the resulting difference equation are then never
evaluated at $p=0$ and no divergence arises. But as this avoidance of
a singularity would merely be put in by hand by choosing a suitable
lattice such a resolution would hardly seem satisfactory. (A more
refined version of this possibility can be found in
\cite{ConsistDisc,ConsistCosmo}.)

\begin{figure}
  \includegraphics[height=.15\textheight]{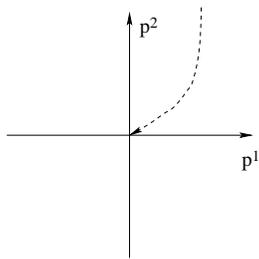}
  \caption{Anisotropic minisuperspace with a classical trajectory and
    the singular hyperplanes.
    \label{AnisoCross}}
\end{figure}

As we will see later, quantizations which pass the bounded isotropic
curvature test of quantum hyperbolicity do exist.  Bounded curvature
in isotropic models thus provides a simpler but non-trivial test of
quantum hyperbolicity. It is, however, not to be over-generalized, as
it has occasionally been done in the recent literature, as anisotropic
models show.  In anisotropic models the classical singularity occurs
at $p^1=p^2=p^3=0$. This is a single point in the interior of
3-dimensional minisuperspace. But unlike the classical trajectory
sketched in Fig.~\ref{AnisoCross}, a wave function will be supported
on 3-dimensional regions.  Coefficients of the dynamical equations
then can become singular even if a single $p^I$ vanishes. All three
planes $p^I=0$ are thus to be considered potentially singular for
quantum dynamics.

As before, coefficients of the diff.\ equation must not diverge for a
well-posed initial value problem. Again, the danger comes from a
matter Hamiltonian but possibly also from intrinsic curvature which can
diverge in anisotropic models.  Now, however, {\em boundedness is not
  required} on the whole minisuperspace; only a local version in any
finite subset of the singular hyperplanes is necessary.  Curvature can
(and usually does) grow parallel to singular submanifolds, e.g.\ for
$p^1\to\infty$ while $p^2$ and $p^3$ remain small
\cite{Spin,NonChaos,ChaosLQC}. This shows that bounded curvature
operators are not required in any non-isotropic model. An analysis of
the singularity issue then has to refer to more detailed properties of
quantum dynamics and the corresponding initial value problem.

This is also the case in any inhomogeneous model. In such cases, many
coupled diff.\ equations result, one for each spatial point as the
equations are functional. Although equations remain linear, initial or
boundary value problems are much more complicated and to be analyzed
in detail.  Without any symmetry it would not even be known how to
locate classical singularities on superspace in a one-to-one manner.
This complicates any scheme to resolve singularities, not just the
principle of quantum hyperbolicity. Moreover, in full situations one
may have to expect non-commutative metric operators to occur (see,
e.g., \cite{NonCommFlux}). If singularities are identified using
metric components, such a non-commutativity would wash out classical
results \cite{DegFull}. In such situations, only general principles
can be applied which is reminiscent of the very general appearance of
classical singularities in singularity theorems of general relativity.
But even before such general situations can be addressed, in a
combination of mathematical relativity with quantum dynamics, many
non-trivial tests of quantum gravity arise from several symmetric or
other models.  The main criterion in those cases is whether or not the
resulting initial value problem with given quantum recurrence
relations is well-defined.

\section{Canonical Quantization}

To address the singularity issue from the point of view of quantum
hyperbolicity, i.e.\ the unique extendability of states across
classical singularities, a sufficiently detailed formulation of
quantum dynamics is needed.  Proper wave functions are required,
independent of any background metric. Only then can quantum geometry
be taken into account fully which on smaller scales or close to a
classical singularity can be very different from classical geometry.
Canonical quantizations allow systematic constructions of quantum
field theories even without reference to a background metric. The
metric itself can then be a full quantum operator and, e.g.,
fluctuate. Its quantum dynamics is relevant for the singularity issue.

In canonical gravity, there is an infinite dimensional phase space, in
ADM variables \cite{ADM}, of fields $q_{ab}$, the spatial metric, and
momenta $p^{ab}$, related to extrinsic curvature. The other components
$N$ and $N^a$ of the space-time metric (\ref{CanonMetric}) are not
dynamical since $\dot{N}$ and $\dot{N}^a$ do not occur in the action.
Thus, according to the usual definition as the derivative of the
action by time derivatives of fields, momenta $p_N$ and $p_{N^a}$
vanish identically. Rather than determining the evolution of $N$ and
$N^a$, their equations of motion $0=\dot{p}_N=-\frac{\delta H}{\delta
N}$ and $0=\dot{p}_{N^a}=-\frac{\delta H}{\delta N^a}$ imply
constraints on the phase space variables: the Hamiltonian constraint
\[
 0=\frac{\delta H}{\delta
N}=\frac{\sqrt{\det q}}{16\pi G}{}^{(3)}\!R- \frac{16\pi
G}{\sqrt{\det q}} (p_{ab}p^{ab}-{\textstyle\frac{1}{2}}(p^a_a)^2)
\]
 and the diffeomorphism constraint
\[
 0=\frac{\delta H}{\delta N^a}= 2D_bp_a^b\,.
\]

For a generally covariant theory the total Hamiltonian is a sum of
constraints since no preferred time variable exists to which absolute
evolution generated by a non-vanishing Hamiltonian would refer. For
general relativity, we have the Hamiltonian
\begin{equation} \label{totalH}
 H=H[N]+D[N^a]=\int\md^3x N(x)\frac{\delta H}{\delta
N}+\int\md^3x N^a(x)\frac{\delta H}{\delta N^a}
\end{equation}
and the constraints determine the full dynamics. They constrain
allowed values of the fields and their initial data, and they generate
Hamiltonian equations of motion in coordinate time through Poisson
brackets such as $\dot{q}_{ab}=\{q_{ab},H\}$. On the right hand side
of those equations of motion lapse and shift occur through
(\ref{totalH}).  For specific equations of motion, they thus have to
be specified by choosing a space-time gauge. This determines which
coordinate the dot in equations of motion refers to (e.g.\ proper or
conformal time depending on whether $N=1$ or $N=a$ in isotropic
cosmology).

This is different in quantum gravity since there will be no reference
to the space-time coordinates at all. Dynamics must be described in a
gauge-independent manner rather than using space-time coordinates.
While this is also possible, though complicated, to do in classical
gravity, in background independent quantum gravity it is the only
option.  The tensors $q_{ab}$ and $p^{ab}$ are then to be replaced by
operators, acting on states such as $\psi[q_{ab}]$ solving the
infinitely many quantum constraints
\[
 \hat{H}[N]\psi=\hat{D}[N^a]\psi=0 \quad \mbox{for all }N\mbox{ and
}N^a\,.
\]
The difficult part is to define precisely which function space
``$\psi[q_{ab}]$'' refers to, how basic operators quantizing $q_{ab}$ and
$p^{ab}$ are represented in a well-defined way, and how their products
or even non-polynomial expressions in the constraints are being dealt
with.

All these steps simplify in homogeneous models such as in isotropic
Wheeler--DeWitt cosmology. Classically, $q_{ab}= a^2\delta_{ab}$ just
requires a single variable $a$ to be quantized without any tensor
transformation laws to be taken care of in this restricted class of
coordinates respecting the symmetry.\footnote{In the spatially
  flat case the scale factor can be rescaled by a constant, which does
  however not introduce coordinate dependent pre-factors.} Wave
functions $\psi(a)$ are simply square integrable functions of a single
variable. The Wheeler--DeWitt equation is obtained by quantizing the
Hamiltonian constraint:
\[
 \frac{1}{a}\frac{\partial}{\partial a}\frac{1}{a}
  \frac{\partial}{\partial a}a\psi\propto \hat{H}_{\rm matter}\psi
\]
where there is some freedom in choosing the operator ordering of $a$
and $\partial/\partial a$ (the ordering here agrees with (\ref{WdW})).
The diffeomorphism constraint $\hat{D}=0$ vanishes identically.

\subsection{Index-free objects}

In general, a well-defined background independent quantization is much
more complicated due to, for instance, non-trivial transformation
properties of tensorial basic variables.  A single component
$q_{\mu\nu}$ does not have coordinate independent meaning, but only
the tensor $q_{ab}$ has. If we quantize single components, the
question is which coordinate system an operator $\hat{q}_{\mu\nu}$
should refer to. Properties of space are to be determined by states
(e.g.\ through expectation values) only after operators have been
defined. The space-time manifold in general relativity is part of the
solution to Einstein's equation, whose coordinates follow from
coordinates on an initial spatial slice once the gauge has been fixed
fully. Thus, it becomes available only after the classical constraint
and evolution equations have been solved. In quantum gravity, we have
to turn the basic tensors given by the spatial metric and its momentum
into operators before we can even formulate the constraint equations.
There is thus no meaning whatsoever to an operator
``$\hat{q}_{\mu\nu}$'' because its classical analog $q_{\mu\nu}$, when
defined in one chosen coordinate system, will receive factors
$\partial x'{}^{\mu'}/\partial x^{\mu}$ when transformed to other
coordinates.  (These coordinates are only spatial because a canonical
formulation deals with spatial tensors. Nevertheless, the classical
tensors transform in this manner on any spatial slice, not just the
``initial'' one.)  One can certainly avoid this by choosing a fixed
set of spatial coordinates on any slice once and for all. But the
resulting quantum theory would keep a trace of that choice and would
be badly non-covariant.

For this reason, no systematic quantization is known in ADM variables
using the spatial metric and its momentum.  No full quantum theory has
been formulated in those variables but one has constructed exclusively
models where one can reduce the metric variables to scalar quantities.
Scalar quantities can then more easily be quantized since they do not
receive coordinate dependent factors from the tensor transformation
law. Examples include homogeneous models where the metric is
determined by a finite number of spatial constants which do not
transform under the allowed coordinate changes preserving the
symmetry. (In fact, even such simple models can give rise to confusion
from coordinate changes. An example is the scale factor of flat
isotropic models which can be rescaled arbitrarily by a constant.
This rescaling freedom cannot be appropriately dealt with in a
quantization of the model unless the freedom is fixed from the
outset.) Also in inhomogeneous models such as Einstein--Rosen or Gowdy
models one can sometimes express the metric in terms of scalar fields
on a given manifold \cite{K:EinsteinRosen,GowdyQuadratic}. But those
are special properties not available in a similar form in full
generality. It then remains unclear if those quantum models show
typical properties of quantum gravity or special features used in
their formulation. In particular, there is no well-defined relation
between those symmetric quantum models and a potential full theory.

To proceed, we thus need to reformulate unrestricted spatial
geometries in terms of index-free objects.  A successful classical
reformulation starts by first removing one spatial index from $q_{ab}$
and $p^{ab}$ (or $K_{ab}$). As already used, we introduce the co-triad
$e_a^i$ such that $e_a^ie_b^i=q_{ab}$. Here, $i$ is just an index
enumerating co-vectors and does not imply transformation properties
under changes of coordinates. Moreover, the position of this index is
irrelevant and will be summed over even when repeated in the same
position.  Similarly, we trade in an index $i$ for a spatial index in
extrinsic curvature by defining $K_a^i:=e^b_iK_{ab}$, using the triad,
i.e.\ the inverse $e^b_i$ of $e_a^j$. We cannot proceed further in
this way and remove all spatial indices because
$e^a_je_a^i=\delta_j^i$ would loose all information about the metric.
The triad, on the other hand, has all information about the metric. In
fact, it has more freedom because an SO(3) rotation $R^j_ie_a^i$
leaves the metric $q_{ab}$ unchanged.  This corresponds to the three
additional components which the (non-symmetric) matrix $e_a^i$ has
compared to the symmetric $q_{ab}$.  The new degrees of freedom are
removed from the resulting field theory by imposing a further
constraint on triad variables which has the form of the Gauss
constraint in an SU(2) Yang--Mills theory.

Using $K_a^i$ as one of the canonical variables leads to a momentum
which is not exactly the triad but its densitized version, the
densitized triad $E^a_i=|\det e_b^j|e^a_i$. At this point we should
recall that densitized triad variables were very convenient for the
singularity issue because they implied positions of classical
singularities in the interior of the space of geometries. At this
point we see how the choice of basic variables also plays a role when
defining a quantum theory. As we proceed, we will see that a
reformulation of classical gravity in terms of index-free objects is
possible precisely in terms of variables which use the densitized
triad instead of the spatial metric or co-triad.

To proceed with the definition of index-free objects, we will replace
the tensor $K_{ab}$ by a connection. This will
immediately suggest an index-free object, a holonomy, and has the
additional advantage that spaces of connections are much better
understood mathematically than spaces of extrinsic curvature tensors
or metrics.  Just as the metric defines a Christoffel connection, the
co-triad defines the spin connection
\begin{equation}\label{SpinConn}
 \Gamma_a^i= -\epsilon^{ijk}e^b_j (\partial_{[a}e_{b]}^k+
 {\textstyle\frac{1}{2}} e_k^ce_a^l\partial_{[c}e_{b]}^l)\,.
\end{equation}
As a functional of $e_a^i$, it has vanishing Poisson brackets with
$E^a_i$ and can thus not be used as a momentum replacing $K_a^i$. But
the Ashtekar connection \cite{AshVar,AshVarReell}
$A_a^i=\Gamma_a^i-\gamma K_a^i$ transforms as a connection, as any sum
of a connection and a tensor does, and is canonically conjugate to the
densitized triad because $K_a^i$ is. In the definition, $\gamma>0$ is
the Barbero--Immirzi parameter \cite{AshVarReell,Immirzi} which does
not have any effect in the classical theory but is important in
quantum gravity.

Loop quantum gravity is based on a canonical quantization of the phase
space spanned by $A_a^i$ and $E^a_i$ with
\begin{equation} \label{Poisson}
 \{A_a^i(x),E^b_j(y)\}=8\pi\gamma G\delta_a^b\delta^i_j\delta(x,y)\,.
\end{equation}
Its success relies on the fact that connections and densitized vector
fields can easily be expressed in terms of index-free objects which
can then be represented on a Hilbert space.  Instead of connection
components we use holonomies
\begin{equation}
 h_e(A)={\cal P}\exp\int_e A_a^i\tau_i\dot{e}^a\md t
\end{equation}
for curves $e$ with tangent vector $\dot{e}^a$ and Pauli matrices
$\tau_j=-\frac{i}{2}\sigma_j$. As usually, ${\cal P}$ denotes path
ordering along $e$ of the non-commuting functions in the integrand. If
holonomies are known for all curves $e$ in space, the connection can
be reproduced uniquely \cite{Giles}.  Similarly, we use fluxes
\begin{equation}
 F_S(E)=\int_S\md^2y n_aE^a_i \tau_i
\end{equation}
for 2-surfaces $S$ with co-normal $n_a=\frac{1}{2}\epsilon_{abc}
\epsilon^{uv}\frac{\partial x^b}{\partial y^u} \frac{\partial
  x^c}{\partial y^v}$ in a parameterization $S\colon y\mapsto x(y)$.
Again, if fluxes are known for all surfaces $S$ in space, the
densitized triad is reproduced uniquely.

Notice that no background metric is used in these definitions as the
tangent vector as well as co-normal are defined intrinsically without
reference to a metric. Moreover, there are no free spatial indices and
the objects transform trivially under changes of coordinates. Instead,
there is a representation of active spatial diffeomorphisms $\phi$
which move along the labels $e\mapsto\phi(e)$ and $S\mapsto\phi(S)$ of
holonomies and fluxes. These objects can thus be represented on a
Hilbert space without having to include coordinate factors in the
tensor transformation law.

\subsection{Loop quantum gravity}

For a representation one also has to know the Poisson algebra of basic
variables which is to be turned into a commutator algebra. Poisson
relations of holonomies and fluxes define the holonomy-flux algebra in
which no delta functions occur even though we are dealing with a field
theory: integrating connections and densitized triads to obtain
index-free objects has automatically introduced the correct kind of
smearing. Any delta function present in the Poisson relation
(\ref{Poisson}) between $A_a^i$ and $E^b_j$ is integrated out in the
holonomy-flux algebra. A well-defined algebra results, ready to be
represented for a quantization.  This provides the kinematical setting
of loop quantum gravity.

There are then two routes for applications of this basic structure: it
allows a formulation of quantum dynamics in a background independent
manner which we are going to describe now. Moreover, the unique
representation of the full theory induces representations of symmetric
sectors derived from a full framework. This leads to symmetric models
with a precise relation to the full theory, which was lacking in
Wheeler--DeWitt quantizations. As we will see in the next section, the
induced representation is inequivalent to the Wheeler--DeWitt
representation even in the simplest models, with important
implications for the singularity issue.

\subsubsection{Representation}

One can now construct a quantum representation of the smeared basic
fields which arose naturally in providing variables suitable for a
background independent quantization. After the basic objects have been
represented, thus providing a quantum theory, one can quantize and
impose constraints as operators. This will ensure that only space-time
covariant observables arise. From the form of the basic variables, an
SU(2) connection and a densitized momentum field in Ashtekar
variables, general relativity appears as a gauge theory subject to
additional constraints.

A representation is most easily constructed in the connection
representation where states are functionals on the space of
connections. This would not be available had we not been led to
introduce connections rather than tensors in constructing index-free
objects; the representation will thus carry characteristic features as
traces of the background independent quantization. In a connection
representation, holonomies act as multiplication operators. Starting
from a basic state which, as a function of connections, is constant,
we thus ``create'' non-trivial states which depend on the connection
along the edges used in multiplicative holonomies. Although such
states depend on the connection only along edges, the resulting states
can be complicated with edges non-trivially being knotted and linked
with each other.  Moreover, edges can intersect each other giving rise
to vertices in which more than two edges meet. The resulting space of
states can be spanned by a basis of spin network states
\cite{RS:Spinnet} defined as
\[
 f_{g,j,C}(A)=\prod_{v\in g} C_v\cdot \prod_{e\in g}
\rho_{j_e}(h_e(A))
\]
where $g$ is an oriented graph collecting all the edges used for
holonomies, with labels $j_e$ indicating irreducible SU(2)
representations $\rho_{j_e}$ in which edge holonomies are evaluated,
and matrices $C_v$ which ensure that matrix elements of holonomies are
multiplied with each other in a way resulting in a gauge invariant
complex valued function of the connection. Since holonomies are
multiplied with SU(2) group elements at the endpoints of their edges
under a gauge transformation, the contraction matrices sit in vertices
where different edges meet each other.

\begin{figure}
  \includegraphics[height=.15\textheight]{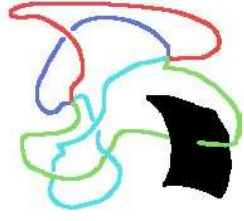}
  \caption{A colored graph representing a spin network state together
    with a 2-surface intersecting a link.
    \label{SpinNetSurf}}
\end{figure}
 
A well-known example of a gauge invariant function of holonomies is
the Wilson loop obtained by taking the trace of a holonomy around a
closed loop. In the fundamental representation, we can reproduce this
in the above language by introducing two vertices $v_1$ and $v_2$
along the loop which split the loop into two non-selfintersecting
edges $e_1$ and $e_2$. We orient this graph such that both edges start
in $v_1$ and end in $v_2$. Their holonomies in the fundamental SU(2)
representation are $2\times2$ matrices $h_I(A)^B_C$ which change under
a local gauge transformation to $g(v_1)^B_Dh_I(A)^D_Eg^{-1}(v_2)^E_C$
where $g(v_I)\in{\rm SU}(2)$ are the values the local gauge
transformation takes in the vertices. There are thus two factors in
each vertex, such as $g(v_1)^{B_1}_{D_1}g(v_1)^{B_2}_{D_2}$ from the
two holonomies, which in the final expression must cancel each other
when contracted with the matrix $C_{v_1}$. Using the identity
$\epsilon_{AB}g^B_C=(g^{-1})^B_A\epsilon_{BC}$ satisfied for any
SU(2)-matrix $g$, one can see that a matrix
$C(v_1)_{AB}=\epsilon_{AB}$ and similarly $C(v_2)^{AB}=\epsilon^{AB}$
results in a gauge invariant function:
$C(v_1)_{AB}g(v_1)^A_Cg(v_1)^B_D=C(v_1)_{CD}$. The resulting spin
network state
\[
 W_g(A) = \epsilon_{AC}\epsilon^{BD}\cdot (h_1(A))^A_B 
(h_2(A))^C_D
 = (h_1(A))^A_B (h_2(A)^{-1})^B_A
 = {\rm tr}(h_1(A) h_2(A)^{-1})
\]
leads to the usual expression for the Wilson loop, where in the second
step we used the same SU(2) identity as before. At intersection points
of higher valence, one can use decompositions of tensor products of
SU(2) representations into irreducible ones, i.e.\ the usual
recoupling rules known from angular momentum in quantum mechanics, to
find all vertex matrices $C_v$ leading to gauge invariant results.

Spin network states span the whole quantum representation space
because the action of holonomies as basic configuration variables is
complete. On any such state, holonomies are represented as
multiplication operators as used in the construction of the states.
Since these operators have to respect unitarity properties
representing the classical reality conditions of $A_a^i$, an inner
product of the representation space results. In this process spin
network states turn out to be an orthonormal basis provided one
chooses an orthonormal representation of the vertex matrices $C_v$
(which for a given number of edges meeting the vertex and given edge
labels form only a finite-dimensional space as it follows from the
rules of tensor decomposition). When completed, this defines the
kinematical Hilbert space of loop quantum gravity.

Fluxes are conjugate to holonomies and thus become derivative
operators on the Hilbert space. Using
\begin{equation}
 \hat{F}_S f_g = -8\pi i\gamma G\hbar \int_S {\rm d}^2y\tau^i n_a
\frac{\delta}{\delta A_a^i(y)}f_g(h(A))
 =-i\gamma\ell_{\rm P}^2 \sum_{e\in g}\int_S {\rm d}^2y\tau^i n_a
\frac{\delta (h_e)^A_B}{\delta    A_a^i(y)}
\frac{\partial f_g(h)}{\partial (h_e)^A_B}\,,
\end{equation}
which is a consequence of the chain rule for $A_a^i$ derivatives
acting on functions of holomies, shows that non-zero contributions
result only if $S$ intersects the edges of $g$. Indeed, the functional
derivative of holonomies (assuming, to be specific, that the surface
intersects the edge at its starting point) gives
\[
  \int_S\md^2yn_a\frac{\delta h_e}{\delta
  A_a^i(y)}=\frac{1}{2}\tau_i
 \int_S\md^2y\int_e\md t
  n_a(y) \dot{e}^a \delta(e(t),y) h_e= \frac{1}{2}{\rm Int}(S,e)\tau_ih_e
\]
and thus a factor ${\rm Int}(S,e)$ of the oriented intersection number
of the surface $S$ and an edge $e$.  Individual contributions in the
sum over intersection points are then determined by ``angular momentum
operators'' (su(2) derivatives) $(\tau_ih)^A_B\partial/\partial h^A_B=
{\rm tr}(\tau_ih\partial/\partial h)$ acting on holonomies. Since
these operators have discrete spectra and are summed over in an at
most countable sum, all flux operators have discrete spectra. With
fluxes being the basic operators representing spatial geometry through
the densitized triad, discrete spatial geometry emerges from the
construction without being put in in the first place.

This representation is not only convenient to construct and to work
with, it is also, under mild assumptions, the unique representation of
the algebra of holonomies $h_e$ and fluxes $F_S$ on which the
diffeomorphism group acts unitarily \cite{LOST,WeylRep,WeylRepII}.
Classically, diffeomorphisms $\phi$ act on holonomies and fluxes by
moving the defining submanifolds, $(h_e,F_S)\mapsto
(h_{\phi(e)},F_{\phi(S)})$.  If this is required to carry over
to the quantum theory, as it should be since any violation of
unitarity of the diffeomorphism group would imply a breakdown of
spatial background independence, no other representation is possible.
As it happens often, the requirement of a symmetry reduces the class
of available representations. With the large diffeomorphism group as a
consequence of background independence, the representation appears to
be selected uniquely.

\subsubsection{Quantum geometry}

While fluxes do not have direct intuitive implications for spatial
geometry, they occur in more typical objects such as the area
operator. The area of a surface $S$ with co-normal $n_a$ as used in
the definition of fluxes is $A(S)=\int_S \md^2y \sqrt{E^a_in_a
  E^b_in_b}$. A quantization thus requires a product of flux operators
which can be defined after regularization \cite{AreaVol,Area}. Due to
the square of triad components present in the classical expression,
the quantum operator contains a square of angular momentum operators
whose spectrum is well-known. This allows one to determine the area
spectrum
\[
 \hat{A}(S)f_{g,j} = \frac{1}{2}\gamma\ell_{\rm P}^2\sum_{p\in S\cap g}
 \sqrt{j_p(j_p+1)} f_{g,j}
\]
valid for the case where
no intersections between $S$ and the graph occur in vertices of
$g$. In the general case the spectrum is more complicated but also
known explicitly.

Similarly, a volume operator $\hat{V}$ is obtained by quantizing the
classical expression $V(R)=\int_R\md^3x \sqrt{|\det E|}$. Again after
regularization, contributions now come only from vertices $v\in R\cap
g$ whose values are constructed from the invariant matrices in the
vertex \cite{AreaVol,Vol2}. Although the spectrum is much more
complicated to determine than the area spectrum and not known
completely, it is discrete.

The volume spectrum contains zero as a highly degenerate eigenvalue
which is realized for instance, but not exclusively, in the case where
no vertex lies in the region $R$. But also the total volume of the
whole spatial slice has a highly degenerate zero eigenvalue even for
vertices with arbitrarily high valence. Therefore, there is no densely
defined inverse of $\hat{V}$. However, when we come to matter
Hamiltonians we will need the inverse determinant of $E^a_i$, such as
in the case
\[
  H_{\varphi}=\int\mathrm{d}^3x\left( \frac{1}{2}
   \frac{p_{\varphi}^2+E^a_iE^b_i\partial_a\varphi
     \partial_b\varphi}{\sqrt{|\det E^c_j|}}+\sqrt{|\det
     E^c_j|}V(\varphi)\right)
\]
of a scalar field.  Since also metric components entering matter
Hamiltonians have to be quantized in quantum gravity, we need an
inverse volume operator. As the spatial volume vanishes usually at
classical singularities, this issue is related to the singularity
problem. In fact, in the context of quantum hyperbolicity we have
already seen that such coefficients matter for the well-posedness of
initial value problems of wave functions.

\subsubsection{Quantization and ambiguities}

Using
\begin{equation} \label{ident}
 \left\{A_a^i,\int\sqrt{|\det E|}\mathrm{d}^3x\right\}= 2\pi\gamma G
 \epsilon^{ijk}\epsilon_{abc} \frac{E^b_jE^c_k}{\sqrt{|\det E|}}
\end{equation}
and approximating $A_a^i$ by holonomies one can replace inverse powers
of $\det E$ by positive powers in Poisson brackets \cite{QSDI}.
Inserting holonomies and the volume operator, and replacing the
Poisson bracket by $(i\hbar)^{-1}$ times a commutator then results in
well-defined operators with a classical limit as required.  But the
resulting operators are not identical to an inverse volume which does
not exist in the quantum theory. Deviations between the classical
inverse and the quantum behavior thus result which are most noticeable
at small length scales.

With many different ways to re-write, e.g.\ using different
representations for holonomies, giving the same classical expression
but differing in quantum properties such as their spectra,
the operators are subject to quantization ambiguities. As they are
non-basic operators and in fact non-polynomial functions of the basic
ones, this is not surprising and would occur in any quantum theory.

\subsubsection{Dynamics}

Also the constraints, in particular the Hamiltonian constraint
\begin{equation}
 H[N] = \frac{1}{16\pi G} \int_{\Sigma} \mathrm{d}^3x N
 \left(\epsilon_{ijk} F_{ab}^i\frac{E^a_jE^b_k}{\sqrt{|\det
E|}} -2(1+\gamma^{-2})
 (A_a^i-\Gamma_a^i)(A_b^j-\Gamma_b^j) \frac{E^{[a}_iE^{b]}_j}{\sqrt{|\det E|}}
 \right)
\end{equation}
written in Ashtekar variables with
$F_{ab}^i=2\partial_{[a}A_{b]}^i+ \epsilon^{ijk}A_a^jA_b^k$,
are non-polynomial functions of the basic variables and thus subject
to quantization ambiguities. But given the difficulties of other
attempts to formulate quantum gravity, even a single well-defined
quantization of the Hamiltonian constraint would be a success. With
the techniques described before, this can be accomplished \cite{RS:Ham,QSDI}.

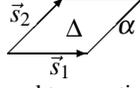
\begin{figure}
\centerline{ \begin{picture}(50,22)(0,0)
\put(0,0){\vector(1,0){30}} \put(0,0){\vector(1,1){21.2}}
\put(30,0){\line(1,1){21.2}} \put(21.2,21.2){\line(1,0){30}}
\put(20,-5){\makebox(0,0){$\vec{s}_1$}}
\put(5,15){\makebox(0,0){$\vec{s}_2$}}
\put(25,10){\makebox(0,0){$\Delta$}}
\put(45,10){\makebox(0,0){$\alpha$}}
\end{picture}
}
\caption{Loop $\alpha$ used to quantize curvature components. \label{loop}}
\end{figure}

Also here, we need an inverse determinant of the triad which follows
from the relation (\ref{ident}).  We have not encountered the
curvature components $F_{ab}^i$ before in operators, but they can be
quantized by using the relation
\[
 s_1^as_2^b F_{ab}^i\tau_i=
\Delta^{-1}(h_{\alpha}-1) +O(\Delta)
\]
used also in lattice gauge theories. Here, $\alpha$ is a square loop
of coordinate size $\Delta$ and with tangent vectors $s_1^a$, $s_2^a$
in a vertex as shown in Fig.~\ref{loop}.  One thus replaces the
curvature components by holonomies around small loops which can then
directly be represented on the Hilbert space. For this, one has to
choose a prescription for the loops since the classical expression is
simply evaluated in a point. The prescription gives rise to further
quantization ambiguities which are, however, not as large as they
would be for matter field theories on a background metric due to
diffeomorphism invariance: only knotting and linking of the loop with
edges in the graph of the state matters but not the precise position
of an embedding in space.

The remaining terms in the constraint involve extrinsic curvature and
are the most complicated to deal with. We would have to subtract the
spin connection from the basic Ashtekar connection to obtain extrinsic
curvature components. Ashtekar connection components could again be
quantized using holonomies, but the spin connection (\ref{SpinConn})
in general is a complicated function of the triad components.
Moreover, it is not covariant, not even tensorial, and thus impossible
to quantize directly.  Instead of following this line it is
conceptually easier to use a further identity \cite{QSDI}
\[
 K_a^i=\gamma^{-1}(A_a^i-\Gamma_a^i)
\propto \left\{A_a^i,\left\{\int\md^3x F_{ab}^i 
\frac{\epsilon^{ijk}E^a_jE^b_k}{\sqrt{|\det
E|}},\int{\sqrt{|\det E|}}\mathrm{d}^3x\right\}\right\}
\]
which allows one to express extrinsic curvature through commutators of
the already quantized first term in the constraint with the volume
operator and holonomies. The result is a highly complicated operator,
but it is well-defined. Moreover, it displays crucial properties and
deviations from the classical behavior which can be studied in models
where the operator simplifies.

\subsubsection{Quantum effects}

The typical properties shown by the construction of quantum
Hamiltonians are results of using holonomies, the basic tenet of
loop quantum gravity:
\begin{enumerate}
\item Quantized inverse powers of triad
components give rise to modified small-scale behavior of 
coefficients. For singularities, this may be related to the issue of
boundedness of coefficients in diff.\ equations as discussed before.
\item Replacing local curvature and connection components by
    holonomies along extended loops implies non-locality as well as
    higher order spatial derivatives. This will be seen later to imply
    difference operators in equations for wave functions.
\end{enumerate}
In suitable semiclassical states, the quantum Hamiltonian must have an
expectation value identical to the classical expression to zeroth
order in $\hbar$.  In any interacting (non-linear) theory, however,
there will be quantum corrections which one can formulate in effective
classical equations as will be discussed in the final section. In deep
quantum regimes, full quantum equations have to be used with
potentially very different properties compared to classical behavior
or the Wheeler--DeWitt quantization as we will study them in the next
section.  This provides a general framework for quantum dynamics in
which quantum hyperbolicity is testable.

\section{Loop Quantum Cosmology}

Loop quantum gravity provides a non-perturbative and background
independent formulation of quantum gravity. Its main ingredients are a
well-defined representation of basic fields, spatial discreteness and
candidates for quantum dynamics.  A description to study singularities
from the perspective of quantum gravity is thus in principle provided
but, in such a general setting, difficult to apply. There are not just
severe complications from technical as well as conceptual issues
of quantum gravity, but already the classical understanding of
singularities in general is not precise enough even to decide where
one would have to look for resolved singularities in quantum gravity.

As in classical relativity, it is helpful in such a situation to study
in detail explicitly treatable, usually symmetric situations where
general aspects of quantum dynamics can be seen in action. This will
at least give examples for singularity resolution and can suggest
general mechanisms. By looking at different classes of models one then
has a good chance of deciding whether or not such mechanisms are
general or make use of special properties only realized in such
models. Several unsuccessful examples are known, and also one
so far successful scheme to be discussed here.

When using symmetric models in quantum gravity one should be aware of
differences between classical symmetric solutions, which are exact
albeit special solutions of full general relativity, and symmetric
quantum models.  In contrast to classical solutions symmetric quantum
solutions cannot be exact since uncertainty relations are violated.
Both the configuration variables (the densitized triad $E_a^i$) and
their momenta (the Ashtekar connection $A_a^i$) must be symmetric to
ensure a space-time solution respecting the symmetry everywhere.
Non-symmetric modes of all canonical variables are thus zero which is
possible classically but not in the presence of quantum uncertainty
relations. A more general stability analysis is then required which
has been done in a few cases. Alternatively, one can relax symmetries
once models are well-understood and try to approach the full situation
as closely as possible.

An additional aspect of loop quantum gravity makes symmetric models
worthwhile and shows that they can capture essential ingredients of a
full quantization of gravity. Models can in particular illustrate
consequences of quantum representations which have wide implications
not only for the basic variables represented but for any composite
operator constructed from them. The full representation induces
distinguished representations of basic variables (analogously to
holonomies and fluxes) in symmetric sectors which, at a kinematical
level, are thus derived from the full theory \cite{LivRev}.
Since the full representation is unique, the induced representations
of models are distinguished among all representations one could try in
a mini- or midisuperspace quantization.

The induction proceeds by first identifying symmetric states in the
full setting \cite{SymmRed}. This is possible in a distributional
sense, although the underlying discrete structure prevents the
existence of any non-trivial normalizable state invariant under a
continuous symmetry. The induced representation is then derived
through basic full operators fixing these states, determining the
reduced Hilbert space structure as well as the action of reduced basic
operators (see the corresponding sections in
\cite{AnisoPert,SphSymm,InhomLattice}). Composite operators then are
to be constructed from those basic ones within the model following the
same steps as in a full setting. A derivation from full composite
operators would be more complicated and has so far not been attempted.
But for testing properties of the basic loop representation the
sketched procedure is sufficient. After all, dynamics has not been
defined uniquely in the full theory and even an unambiguous derivation
of the Hamiltonian constraint of a model from a full candidate would
thus not result in a unique quantum dynamics. The situation in models
is thus the same as that in the full theory where several candidates
for dynamics formulated on a distinguished representation exist. The
only difference is that models are often treatable explicitly and thus
allow one to check physical consequences of different proposals for
dynamics. In this way, the theory becomes physically testable.

Models allow one to understand the full theory because its
characteristic properties are preserved during the induction
procedure. Most importantly, discreteness of spatial geometry is
realized in an analogous way as we will see soon. Thus, even in
isotropic models the representation is inequivalent to the
Wheeler--DeWitt one which would have a continuous spectrum of the
scale factor as a multiplication operator on the positive real line.
This is the key reason why loop quantum cosmology, as the theory of
symmetric sectors of loop quantum gravity is called, provides new
insights even for the extensively studied field of quantum cosmology.

In addition to the fact that quantum dynamics is often treatable
explicitly in models, there is the added advantage of a much clearer
classical singularity structure. In isotropic models, for instance,
for the usual matter ingredients it suffices to formulate the
condition $a=0$ to select singular states, which is easily done using
the volume operator.  Direct tests of quantum hyperbolicity then
become possible.

\subsection{Isotropic models}

The basic quantities of an isotropic model formulated in Ashtekar
variables comprise one conjugate pair $(c,p)$ with \cite{IsoCosmo,LivRev}
\begin{equation} \label{IsoVars}
 |p|={\textstyle\frac{1}{4}}a^2\quad,\quad
 c={\textstyle\frac{1}{2}}(k+\gamma\dot{a})\,.
\end{equation}
The only difference to metric variables is the fact that $p$ can take
both signs since it describes a densitized triad. The sign of $p$ then
is the intrinsic orientation of space. The parameter $k$ in $c$ is
$k=0$ for a spatially flat model and $k=1$ for positive spatial
curvature. For negative spatial curvature the connection has a
different form not covered here (see \cite{NegCurv}).

The induced quantum representation must be a quantization of this
finite dimensional model. However, it differs from the usual
Schr\"odinger representation one would use in a Wheeler--DeWitt
quantization. A complete set of orthonormal states is given by
\cite{Bohr}
\begin{equation} \label{IsoStates}
 \langle c|\mu\rangle= e^{i\mu c/2}\quad,\quad \mu\in{\mathbb R}
\end{equation}
which already demonstrates the inequivalence to usual quantum
mechanics where plane waves would not be normalizable. Moreover, the
Hilbert space is non-separable, i.e.\ has an uncountable basis.  Basic
operators, obtained by full flux and holonomy operators fixing the
symmetric states (\ref{IsoStates}) interpreted as distributions, act
on these states by
\begin{equation} \label{IsoRep}
 \hat{p}|\mu\rangle= {\textstyle\frac{1}{6}}\gamma\lP^2\mu|\mu\rangle
\quad,\quad
\widehat{e^{i\mu'c/2}}|\mu\rangle= |\mu+\mu'\rangle \,.
\end{equation}
These operators indeed demonstrate the same properties we saw in the
full theory: $\hat{p}$ has a discrete spectrum since all its
eigenstates $|\mu\rangle$ are normalizable, and $e^{i\mu c/2}$ is
represented but not $c$ itself which is the hallmark of any loop
representation. (It is not possible to derive a $c$-operator by taking
a derivative by $\mu'$ because the matrix elements of
$\widehat{e^{i\mu'c/2}}$ in the basis states are not differentiable.)
There is only a difference in how the discreteness of the spectrum of
geometrical operators such as $\hat{p}$ is realized since the set of
all eigenvalues of $\hat{p}$ is the whole real line.  Nevertheless,
the operator has a discrete spectrum since all its eigenstates are
normalizable. In usual quantum mechanics, with a separable Hilbert
space, this would imply that the set of eigenvalues is a discrete
subset of the real line. On a non-separable Hilbert space as we have
it here with an uncountable basis labeled by $\mu\in{\mathbb R}$,
however, an operator with a discrete spectrum can have any eigenvalue
set. We will later see that the mathematical definition using
normalizability of eigenstates is the one that is also relevant for
the singularity issue. It is thus crucial that this property, rather
than any statement about the set of eigenvalues, is preserved. This is
a consequence of strong restrictions in the full theory from
background independence and the subsequent transfer of the
representation to symmetric models.

\subsubsection{Difference equation}

With the induced representation, we can follow most of the steps done
in full to construct Hamiltonian constraint operators. The classical
Hamiltonian constraint with contributions from a matter Hamiltonian
$H_{\rm matter}$ is
\begin{equation} \label{IsoHam}
 H=-\frac{3}{8\pi G}\left[\gamma^{-2}(c-k/2)^2+k^2/4\right]\sqrt{|p|}+H_{\rm
matter}(p,\varphi,p_{\varphi})=0
\end{equation}
which can easily be seen to reduce to the Friedmann equation (with
energy density $\rho=|p|^{-3/2}H_{\rm matter}$) once $c$ and $p$ are
replaced in terms of $a$ and $\dot{a}$ using (\ref{IsoVars}). This
expression needs to be quantized by using ``holonomies''
$e^{i\mu_0c/2}$, with some $\mu_0\in{\mathbb R}$ to be chosen, for the
connection components and Poisson brackets for the triad components.
(Although the latter part is not necessary since there is no inverse
of $p$ due to cancellations from isotropy, we keep this step in order
to have the quantization as close to the full one as possible.) The
action of the resulting operator\footnote{There are different versions
  of this operator in the literature, which partially reflects the
  freedom existing in the full theory as well as details of the
  reduction to isotropic models which have not fully been worked out
  yet. For instance, one may reorder the operator, the one written
  here being non-symmetric, or include several effects which may be
  expected in an inhomogeneous quantization \cite{InhomLattice,APSII}.
  We will mention these possibilities here only when they are relevant
  for the singularity issue.} can be determined explicitly, choosing
$\mu_0=1$:
\begin{equation} \label{IsoHamOp}
 (\hat{H}-\hat{H}_{\rm matter})|\mu\rangle =
 \frac{3}{16\pi G\gamma^3\ell_{\mathrm{P}}^2}
 (V_{\mu+1}-V_{\mu-1})
 (e^{-ik}|\mu+4\rangle
 -(2+k^2\gamma^2)|\mu\rangle+
 e^{ik}|\mu-4\rangle)
\end{equation}
with the quantized matter Hamiltonian $\hat{H}_{\rm matter}$.
In semiclassical regimes extrinsic curvature is small, $c-k/2\ll1$,
which by construction leads to the correct classical limit even though
exponentials of $c$ have been used.

The operator equation $\hat{H}|\psi\rangle=0$ to be solved for
physical states can be expressed as a set of equations for expansion
coefficients $\psi_{\mu}(\varphi)$ in
$|\psi\rangle=\sum_{\mu}\psi_{\mu}(\varphi)|\mu\rangle$ which represent
the state in the triad rather than connection representation.
Applying the operator to such a general state and comparing
coefficients of $|\mu\rangle$ results in the difference equation
\begin{eqnarray} \label{IsoDiffEq}
&&    (V_{\mu+5}-V_{\mu+3})e^{ik}\psi_{\mu+4}(\varphi)- (2+k^2)
(V_{\mu+1}-V_{\mu-1})\psi_{\mu}(\varphi)
+    (V_{\mu-3}-V_{\mu-5})e^{-ik}\psi_{\mu-4}(\varphi)\\
  &=& -{\textstyle\frac{4}{3}}\pi
G\gamma^3\ell_{\rm P}^2\hat{H}_{\rm matter}(\mu)\psi_{\mu}(\varphi) 
\nonumber
\end{eqnarray}
for $\psi_{\mu}(\varphi)$ written in terms of
volume eigenvalues $V_{\mu}=(\gamma\ell_{\rm P}^2|\mu|/6)^{3/2}$
entering through commutators of holonomies with the volume operator.

This defines the dynamical equation for wave functions on
minisuperspace, which can be applied now especially in a neighborhood
of the classical singularity at $\mu=0$ in the interior. There are two
sides to the classical singularity thanks to the triad orientation.
The key question then is whether quantum propagation stops at $\mu=0$
as the classical evolution would at $p=0$.  We have the matter
Hamiltonian in a coefficient of the difference equation, which must be
well-defined at $\mu=0$. This is our first test implied by quantum
hyperbolicity in isotropic models as depicted in Fig.~\ref{IsoCross}.
The matter Hamiltonian $H_{\varphi}=\frac{1}{2}|p|^{-3/2}
p_{\varphi}^2+|p|^{3/2} V(\varphi)$, as before, contains $p^{-1}$
which cannot be quantized directly since $\hat{p}$ has a discrete
spectrum containing zero and thus lacks a densely defined inverse.

\subsubsection{Isotropic curvature bounds}

This is the place where the definition of a discrete spectrum through
the normalizability of eigenstates is key since this, rather than
properties of the set of eigenvalues other than zero being contained
in the set, implies the non-existence of an inverse operator. It thus
seems that the situation is even worse than in a Wheeler--DeWitt
quantization since coefficients in the matter Hamiltonian appear
impossible to define at all. But classically, $a^{-3}$ can be
rewritten in a form suitable for quantization, mimicking the full
identity (\ref{ident}) of \cite{QSDI}, as
\[
 a^{-3}= \left(\frac{3}{8\pi\gamma Glj(j+1)(2
 j+1)}\sum_{I=1}^3{\rm tr}_j(\tau_I
 h_I\{h_I^{-1},|p|^l\})\right)^{3/(2-2l)}
\]
using only positive powers of $p$ and ``holonomies''
$h_I=e^{c\tau_I}$. All this can directly be quantized, with the
Poisson bracket becoming a commutator.  Rewriting in this way
introduces ambiguities because it can be done in many classically
equivalent ways. Making different choices does, however, influence the
quantization. Here, we have indicated two possibilities, taking
different powers of $V$ in the Poisson bracket or taking the trace in
different representations. Eigenvalues of the resulting operators can
be computed explicitly,
\begin{equation}  \label{da}
\widehat{d(a)}_{\mu}^{(j,l)} =
\left(\frac{9}{\gamma\ell_{\rm
P}^2lj(j+1)(2j+1)} \sum_{k=-j}^j
k|p_{\mu+2k}|^l\right)^{3/(2-2l)}
\end{equation}
showing the dependence on the parameters $j$ and $l$.

\begin{figure}
  \includegraphics[height=.2\textheight]{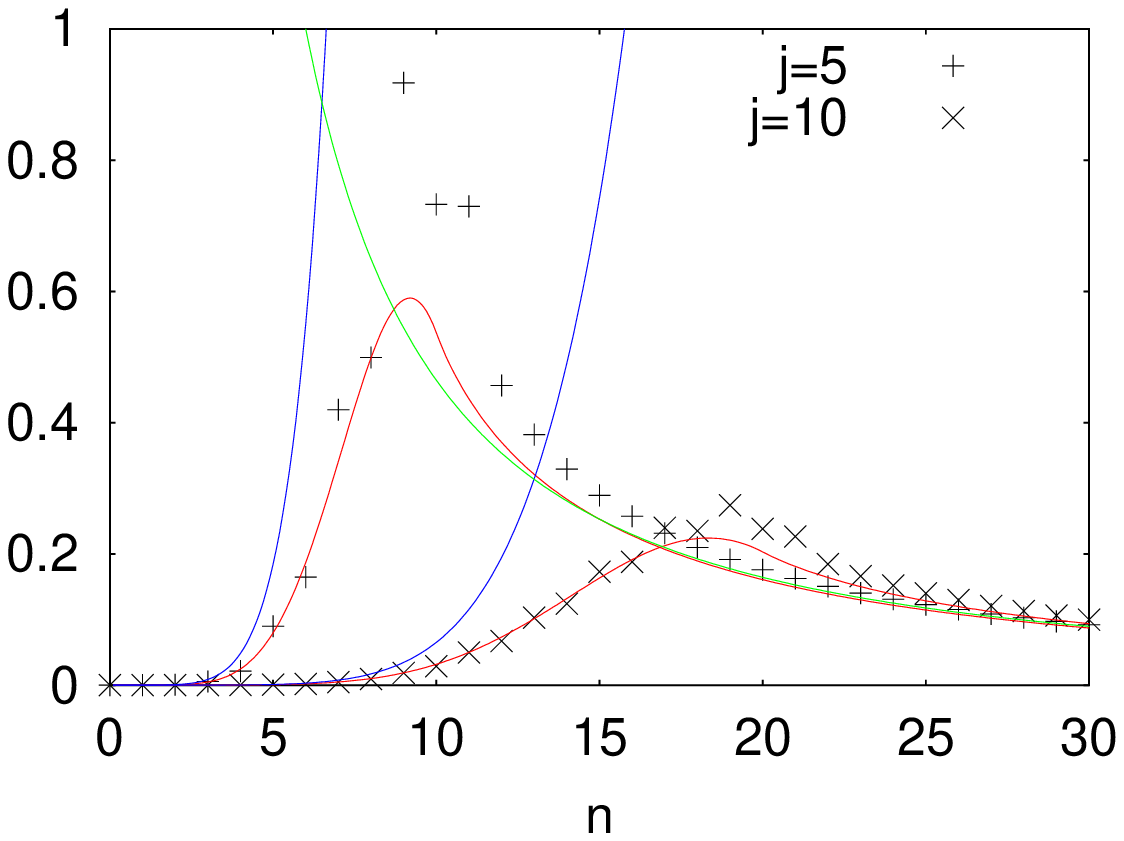} \hspace{1cm}
\includegraphics[height=.2\textheight]{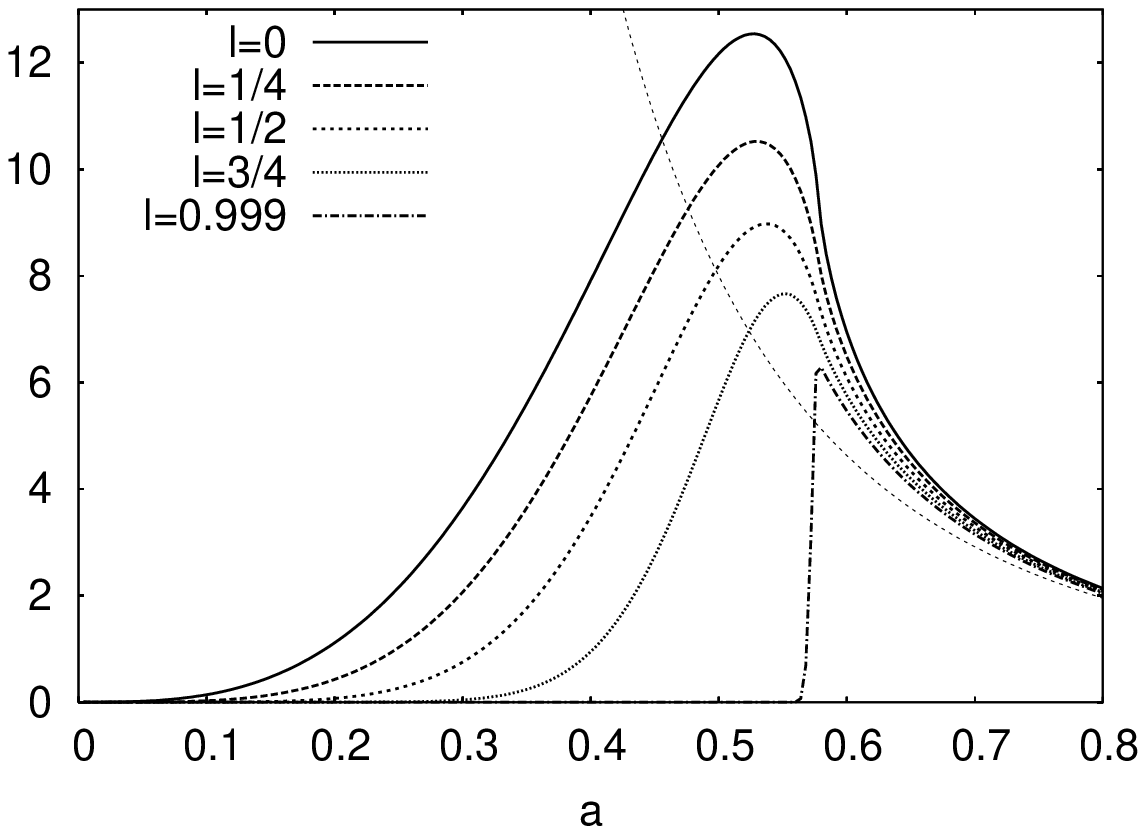}
  \caption{Eigenvalues of the inverse scale factor cubed for two
    values of $j$ and $l=3/4$, together with the effective
    approximation and small-$a$ power laws on the left. (The eigenvalues are
    plotted only for a discrete subset of integer values
    $\mu=n\in{\mathbb N}$.) Effective approximation of the inverse 
    scale factor cubed    for different values of $l$ on the right.
    \label{Dens}}
\end{figure}

Despite of the ambiguities, there are crucial common properties to all
these quantizations. The divergence of the classical $a^{-3}$ is cut
off by quantum effects which is clear from the fact that the operator
is well-defined on all basis states, even on $|0\rangle$ which
corresponds to the classical singularity \cite{InvScale}. The
expression for eigenvalues shows that inverse scale factor operators
in fact all annihilate this state, irrespective of $j$ and $l$, since
for $\mu=0$ one is summing an odd expression in $k$ from $-j$ to $j$.
One can most easily see the behavior around $\mu=0$ by looking at an
approximation of the eigenvalues valid for larger $j$. Viewing the sum
as a Riemann sum of an integral, one obtains \cite{Ambig}
\[
 d(a)^{(j,l)}_{\rm eff}:=
 \widehat{d(a)}_{\mu(a^2)}^{(j,l)}= a^{-3} p_{
 l}(3a^2/ j\gamma\ell_{\rm P}^2)^{3/(2-2l)}
\]
with $\mu(p)=6p/\gamma\ell_{\rm P}^2$ and
\begin{equation}
 p_l(q) =
\frac{3}{2l}q^{1-l}\left( \frac{1}{l+2}
\left((q+1)^{l+2}-|q-1|^{l+2}\right)- \frac{1}{l+1}q
\left((q+1)^{l+1}-{\rm sgn}(q-1)|q-1|^{l+1}\right)\right) \,.
\end{equation}
Important properties, i.e.\ the approach to classical behavior at
large $\mu$, the peak at small values and the approach to zero for
$\mu=0$ are all robust as can be seen from Fig.~\ref{Dens} \cite{Ambig,ICGC}.

\subsubsection{Isotropic quantum hyperbolicity}

The difference equation (\ref{IsoDiffEq}) for $\psi_{\mu}(\varphi)$ can
be used to investigate quantum hyperbolicity in isotropic models. We
have already seen that the coefficients remain well-defined even at
$\mu=0$. But this is only one first test which is necessary for
quantum hyperbolicity to have a chance of being realized. We must,
most importantly, be able to extend any wave function uniquely across
$\mu=0$. For this, the recurrence scheme determined by the difference
equation must be well-defined, i.e.\ we must be able to compute
$\psi_{\mu}$ from the preceding values which themselves have been
computed from some initial values at large $\mu$. By following this
procedure step by step one can see that evolution does continue from
either side of the classical singularity (at $\mu>0$, say) to a new
branch (at $\mu<0$) preceding the big bang at $\mu=0$. This scheme
with two sides to a classical singularity is provided automatically by
ingredients we were forced to assume in loop quantum gravity
\cite{Sing}. It is, in contrast to bounded curvature which was a
condition only in isotropic models, a general scheme which is even
realized in inhomogeneous models.  Since we know the geometrical
meaning of $\sgn(\mu)$ which changes during the transition through the
classical singularity, we can interpret the process as a change in
orientation: the universe turns its inside out. By restricting
variables to metrics one completely misses this possibility of
non-singular behavior, but also the precise form of quantum dynamics
is necessary for a well-defined transition.  Around $\mu=0$,
discreteness manifest in the difference equation is essential and
classical space-time as a smooth manifold dissolves.

The penetration of the classical singularity is thus non-trivial
despite the discreteness: $\mu=0$ is contained in the lattice and not
simply jumped over. Moreover, leading coefficients of the difference
equation may vanish, which could imply a break-down of the recurrence.
In a backward evolution, for instance, we solve recursively for
$\psi_{\mu-4}$ in terms of $\psi_{\mu}$ and $\psi_{\mu+4}$. This is
only possible if $V_{\mu-3}-V_{\mu-5}$, the coefficient of
$\psi_{\mu-4}$ in the difference equation, is nonzero. But it does
vanish for $\mu=4$ since $V_{\mu}$ depends only on $|\mu|$. The value
$\psi_0$ of the wave function at the classical singularity thus
appears to remain undetermined.

Nonetheless, $\psi_{\mu}$ is determined uniquely for all positive and
negative $\mu$: $\psi_0$ just decouples completely. We can follow the
recurrence to negative values of $\mu$. When determining $\psi_{-4}$,
$\psi_0$ seems necessary. But it drops out of the equation because
then the coefficient $V_1-V_{-1}$ vanishes as well as the matter
Hamiltonian which, as a robust property despite of quantization
ambiguities, annihilates the state $|0\rangle$.  The singular state
$|0\rangle$ is then called {\em mantic}\footnote{``{\sc Manto}: I stand
  still, around me circles time.'' Goethe's Faust} with respect to the
given evolution: It plays a passive role in the recurrence scheme of
quantum evolution. In general, a mantic states can be defined as one
at which the recurrence scheme implied by the Hamiltonian constraint
in the triad representation changes its form.

Mantic states have implications not just for the discussion of
recurrence schemes, but they also imply {\em dynamical initial
  conditions} \cite{DynIn,Essay}: Rather than determining $\psi_0$
which completely dropped out of the equations, the equation for
$\mu=4$ implies a linear relation between $\psi_4$ and $\psi_8$ and
thus an additional linear relation between initial values chosen at
large $\mu>0$. There are thus restrictions on initial values not
independent of but implied by the dynamical law, unlike the situation
in every other area of physics. This implication, however, is more
sensitive to the precise form of the constraint. Such conditions would
be weaker for a symmetric ordering of the operator.

\subsection{Anisotropic quantum hyperbolicity}

As isotropic models are always very special, we have to drop symmetry
conditions and see what remains of the observed mechanism of
singularity removal. We first drop one of the isotropy conditions and
look at space-times which are homogenous but have only one rotational
symmetry axis. Such models are interesting for cosmology, but also for
black hole physics since the interior inside the horizon of the
Schwarzschild space-time is of this form.

The densitized triad for the Schwarzschild interior can be written as
\[
E^a_i\tau^i\frac{\partial}{\partial x^a} =
p_c\tau_3\sin\vartheta\frac{\partial}{\partial
x}+p_b\tau_2\sin\vartheta\frac{\partial}{\partial\vartheta}
-p_b\tau_1\frac{\partial}{\partial\varphi}
\]
in spherical coordinates where factors of $\sin\vartheta$ arise due to
the density weight of $E^a_i$. There are now two independent triad
components, $p_c$ and $p_b$. The determinant of the triad is $\det
(E^a_i)=p_c p_b^2$ and its orientation ${\rm sgn} p_c$ is solely determined
by the sign of $p_c$. The sign of $p_b$ is irrelevant, and in fact
there is a residual gauge transformation $p_b\mapsto -p_b$ left after
partially fixing the SU(2)-gauge by requiring the $x$-component of the
triad to point in the su(2)-direction $\tau_3$ as used above.

The triad determines a spatial metric
\[
 \md s^2 = \frac{p_b^2}{|p_c|}\md x^2+ |p_c|\md\Omega^2
\]
whose comparison with the interior Schwarzschild metric ($r<2m$ in the
Schwarzschild coordinate where $r$ becomes time-like and is called $t$
in what follows), $\md s^2= (2m/t-1)\md x^2+t^2\md\Omega^2$, allows us
to identify the classical singularity on minisuperspace $(p_b,p_c)$:
$p_c=0$ at the Schwarzschild singularity while $p_b=0$ is not a
singularity but the horizon.

This model is loop quantized by a representation
$\hat{p}_b|\mu,\nu\rangle = \frac{1}{2}\gamma\lP^2\mu
|\mu,\nu\rangle$, $\hat{p}_c|\mu,\nu\rangle=
\gamma\lP^2\nu|\mu,\nu\rangle$ of basic triad operators acting on
orthonormal states $|\mu,\nu\rangle$ with $\mu,\nu\in{\mathbb R}$,
$\mu\geq0$. As in isotropic models, one can write the Hamiltonian
constraint equation for states as a difference equation 
(using $\psi_{\mu}=\psi_{-\mu}$) \cite{BHInt}
\begin{eqnarray} \label{AnisoDiffEq}
&& 2(C_{\mu+2}\sqrt{|\nu+2|}+C_{\mu}\sqrt{|\nu|}) 
\psi_{\mu+2,\nu+2}- 
2(C_{\mu-2}\sqrt{|\nu+2|}+C_{\mu}\sqrt{|\nu|}) 
\psi_{\mu-2,\nu+2}\nonumber\\
&&+(\sqrt{|\nu+1|}-\sqrt{|\nu-1|})
\left((\mu+2)\psi_{\mu+4,\nu}-
2(1+2\gamma^2)\mu\psi_{\mu,\nu}+
(\mu-2)\psi_{\mu-4,\nu}\right)\nonumber\\
&&+
2(C_{\mu+2}\sqrt{|\nu-2|}+C_{\mu}\sqrt{|\nu|}) 
\psi_{\mu+2,\nu-2}- 
2(C_{\mu-2}\sqrt{|\nu-2|}+C_{\mu}\sqrt{|\nu|}) 
\psi_{\mu-2,\nu-2}
=0
\end{eqnarray}
where $C_{\mu}=|\mu+1/2|-|\mu-1/2|={\rm sgn} \mu$.

By the same procedure as before we conclude that this recurrence is
singularity free which would also be realized if a matter term were
present. Now, however, a more non-trivial test results: there are two
boundaries of a metric minisuperspace, the horizon at $\mu=0$ and the
classical singularity at $\nu=0$.  Only one direction ($\nu$) is to be
extended in densitized triad variables if the singularity is removed.
But there should be no extension through the boundary corresponding to
the horizon because the validity of the homogeneous model breaks down
outside the horizon. There is thus a non-trivial consistency check of
the scheme: evolution through the classical singularity, but not the
horizon must be realized. As one can see from the difference equation,
the extension is provided in just the right manner. By the general
scheme, wave functions are extended only from one orientation of the
triad to another, which provides the new branch at the other side of
the classical singularity. Since orientation is determined by
$\sgn(p_c)$, only the boundary at $p_c=0$ is penetrated, which
corresponds to the classical singularity. The horizon $p_b=0$, on the
other hand, remains a boundary even for quantum evolution. That this
is indeed non-trivial can be seen by trying to reproduce the scheme in
co-triad or triad variables without the density weight. Although there
is a new branch of the opposite orientation, the position of
singularities and horizons in minisuperspace is different and no
natural singularity resolution follows \cite{ModestoConn}

\subsubsection{Beyond the singularity}

In general, solutions to difference equations, especially those of
higher order as encountered in (\ref{IsoDiffEq}) and
(\ref{AnisoDiffEq}), could be wildly oscillating, such as
$\psi_{\mu}=(-1)^{\mu}$ as a solution of
$\psi_{\mu+2}-2\psi_{\mu}+\psi_{\mu-2}=0$. Wave functions could thus
be sensitive to Planck scale physics even in large volume or small
curvature regimes where classical physics should be a good
approximation.  Even if initial values in a semiclassical regime are
chosen to avoid this, oscillations could develop after evolving
through a classical singularity. This would make it difficult to
interpret the new branch as a classical one even when it extends to
large volume. Moreover, oscillating solutions of difference equations
can even be growing exponentially in amplitude and thus dominate any
non-oscillating part.  That this does not happen is a restrictive
stability condition \cite{FundamentalDisc} which happens to be
satisfied automatically in isotropic models as presented above.

For the Schwarzschild interior, mantic states again play an important
role for this issue. The coefficient $\psi_{2,0}$ of the wave function
at $\mu=2$, $\nu=0$ drops out of the highest order term, which implies
the symmetry of solutions under $\nu\mapsto-\nu$ \cite{GenFuncKS}. If
initial values at $\nu>0$ are chosen so as to suppress oscillations on
small scales (giving a so-called {\em pre-classical} wave function),
oscillations will not arise at the other side. This is in accordance
with staticity we know is realized in the outside region of
Schwarzschild: there should be no difference between the past and the
future of the classical singularity. It is again highly non-trivial
how this is realized here through mantic states around the classical
singularity in minisuperspace, although only the interior is used
which itself is not static. But the classical interior dynamics is
determined by the same constraint as the outside, just restricted to
the interior variables.  Although the quantum dynamics used here is
not derived from an inhomogeneous constraint but constructed within
the model mimicking the construction of the full constraint, we arrive
at quantum solutions in accordance with the classical expectation. The
dynamics is responsible for the symmetry of solutions which was not
required at the kinematical level, in agreement with the static
outside as it is determined by the classical constraint.

So far, it is unclear what happens in an inhomogeneous treatment of
the black hole or with matter terms which would imply back-reaction on
the geometry and non-trivial classical dynamics. A possible scenario
concerning the information loss issue, based on several assumptions
but taking into account singularity removal, has been described in
\cite{BHPara}. Although suitable constraint operators and difference
equations are available \cite{SphSymmHam}, they are very complicated
to analyze. Some results are reviewed in \cite{BlackHoles,Puri}.

Oscillations of solutions are also related to the normalizability of
wave functions in an inner product.  States not only have to solve the
difference equation but, as in any quantum theory, must be
normalizable.  This is not always obvious to do when implementing a
quantum constraint which usually changes the inner product on its
solution space.  In our discussion so far, and for most of the rest of
this section, we can safely ignore this issue since we were able to
show that all solutions are uniquely extended across classical
singularities. This must then also be true for the physically
normalizable ones. Here we see that it is important to show that all
solutions, not just ``generic'' ones, are extended.

Nevertheless, for more precise pictures of the evolution through a
classical singularity knowledge of properties of the physical inner
product can be relevant. One method to derive the inner product is
group averaging \cite{Refined} which can be understood as writing
quantum solutions to the constraint $\hat{H}|\psi\rangle=0$ as
 $\delta(\hat{H})|\psi\rangle= \int\md t\exp(it\hat{H})|\psi\rangle$.
This is difficult to compute for the constraints we have to deal with
here (especially due to the absolute value around $\mu$ which prevents
the use of Fourier transformations, see e.g.\ \cite{Bahr} for possible
alternatives).  But since $\hat{H}$ depends on momenta, its
exponential is related to a shift operator in $\mu$. This suggests
that normalizable states are indeed nearly constant on small scales
and do not show strong oscillations.  This justifies the condition of
pre-classicality which has been introduced on intuitive physical
grounds \cite{DynIn}. Oscillations only arise on larger scales where
the matter term becomes important, or on smaller scales when curvature
itself is large.  Since, as we saw, pre-classicality can always be
achieved locally by choosing suitable initial conditions, but may be
difficult to achieve on both sides of a classical singularity, strong
restrictions on the quantization can be expected from a consistent
physical normalizability.

\subsubsection{Unbounded curvature}

\begin{figure}
  \includegraphics[height=.3\textheight]{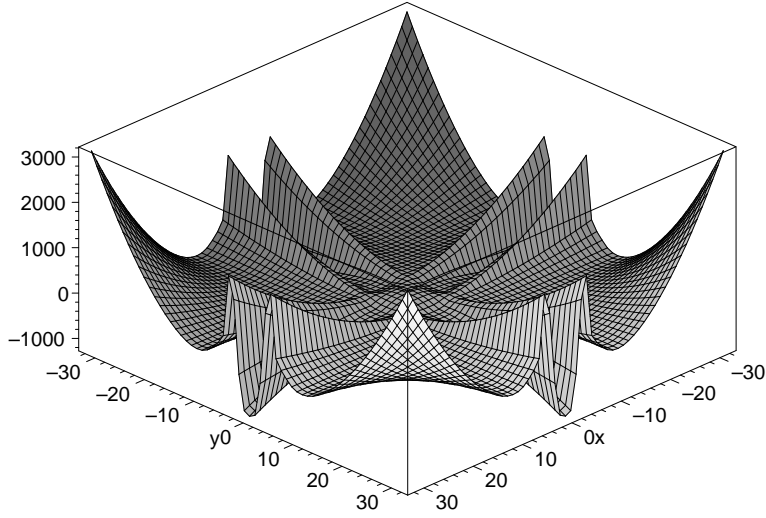}
  \caption{Effective curvature potential in a Bianchi IX model 
\cite{Spin,ChaosLQC,NonChaos}. While it is finite around the isotropy
point in the center, despite of classical divergences in this region,
it is unbounded when large anisotropies are reached.
\label{Pot4}}
\end{figure}

In anisotropic models, intrinsic curvature terms or the matter
Hamiltonian occur in coefficients of the difference equation and must
be well-defined everywhere for a consistent recurrence scheme. As seen
before, in isotropy this implied finite inverse volume if quantum
hyperbolicity is realized, but it is a very special case. Indeed we
have a more general behavior in anisotropic models.  Non-isotropic
quantum hyperbolicity does not require boundedness of curvature in
this sense, and it is indeed not realized generally as can be seen
from Fig.~\ref{Pot4}.

But as we saw, the recurrence relations still do not break down at
classical singularities and quantum hyperbolicity is realized. Bounded
curvature on all of minisuperspace is thus not required for quantum
hyperbolicity. In loop quantum gravity, unboundedness occurs in such a
way that it does not prevent quantum hyperbolicity \cite{DegFull}.

\subsection{Inhomogeneous models}

In inhomogeneous models we have not just one constraint equation but
infinitely many ones since the Hamiltonian constraint has to be
satisfied for any lapse function.
Moreover, these are coupled equations although, for a wave function
$\psi$, they remain linear.
For a spherically symmetric model, we have states
\begin{equation}
 |\psi\rangle=
 \sum_{\vec{k},\vec{\mu}}\psi(\vec{k}, \vec{\mu})
{\unitlength=0.2mm\begin{picture}(200,10)(0,2)
    \put(0,5){\line(1,0){200}} \put(50,5){\circle*{5}}
    \put(100,5){\circle*{5}} \put(150,5){\circle*{5}}
 \put(50,-3){\makebox(0,0){$\mu_-$}}
 \put(100,-3){\makebox(0,0){$\mu$}}
 \put(150,-3){\makebox(0,0){$\mu_+$}}
 \put(25,10){\makebox(0,0){$\cdots$}}
 \put(75,12){\makebox(0,0){$\scriptstyle k_-$}}
 \put(125,12){\makebox(0,0){$\scriptstyle k_+$}}
 \put(175,10){\makebox(0,0){$\cdots$}}
\end{picture}}
\end{equation}
with labels $k_e\in{\mathbb Z}$ for the edges and
$0\leq\mu_v\in{\mathbb R}$ for vertices. Again using a triad
representation, coefficients of the wave function are 
subject to coupled difference equations (one for each edge)
\begin{eqnarray} \label{InhomDiffEq}
 && \hat{C}_0(\vec{k})\psi(\ldots,k_-,k_+,\ldots)+
 \hat{C}_{R+}(\vec{k})\psi(\ldots,k_-,k_+-2,\ldots)
 +\hat{C}_{R-}(\vec{k})\psi(\ldots,k_-,k_++2,\ldots)\nonumber\\
 &&+\hat{C}_{L+}(\vec{k})\psi(\ldots,k_--2,k_+,\ldots)
 +\hat{C}_{L-}(\vec{k})\psi(\ldots,k_-+2,k_+,\ldots)=0
\end{eqnarray}
on an extended superspace of densitized triads. All coefficients
$\hat{C}_I(\vec{k})$ are operators on the vertex labels $\mu$ which we
suppressed in the notation. They have all been computed explicitly in
\cite{SphSymmHam}. Local orientation ${\rm sgn}\det E$ is determined
by ${\rm sgn} k_e$ such that we have to investigate the behavior of
the coupled difference equations around vanishing edge labels.

Again, evolution is non-singular \cite{SphSymmSing} which here depends
crucially on the form (especially possible zeros) of coefficients
$\hat{C}_{R\pm}(\vec{k})$ in a way which is much more non-trivial than
in isotropic models.  Unlike in homogeneous models, a symmetric
ordering is required to extend solutions. Still the solution space is
restricted by dynamical initial conditions as a consequence of mantic
states. This shows that extending models to include more degrees of
freedom does lead to tighter conditions on the allowed quantizations.
So far one scheme is working in all situations considered, a highly
non-trivial result given the complexity of equations such as
(\ref{InhomDiffEq}) compared to the much simpler case of
(\ref{IsoDiffEq}).

\subsection{General properties}

The preceding examples exhaust all types of models where triad
representations exist and which are so far treatable explicitly.  They
allow possible general considerations: assuming singularities of
BKL-type, only homogeneous behavior and diagonal metric or triad
components are essential close to most interesting classical
singularities. This would imply the existence of a triad
representation at least for good approximations also around general
singularities where the above arguments discussed in models would go
through. That neither inhomogeneities nor local degrees of freedom by
themselves spoil quantum hyperbolicity in loop quantum gravity follows
from the demonstration that spherically symmetric and polarized
cylindrically symmetric models respect the mechanism
\cite{SphSymmSing,SphSymmHam}.

If this is not realized, non-commuting triad operators have to be
taken into account. Metric components would thus be unsharp and the
singularity appears washed out. An example realized in loop quantum
gravity, as a consequence of non-Abelian SU(2)-holonomies, is
discussed in \cite{DegFull}. While $\widehat{d(a)}=\widehat{a^{-3}}$
used in (\ref{da}) has the same eigenstates as $\hat{a}$ or the volume
operator $\hat{a}^3$, this is not the case if full SU(2) holonomies
rather than U(1) elements $e^{i\mu c/2}$ are used. Using the basic
equation (\ref{ident}), an inverse power is replaced by
\[
 \dot{e}^a\{A_a^i,V\}\approx 2\epsilon^{-1}{\rm tr}(\tau_i h_e\{h_e^{-1},V\})
\]
with a holonomy $h_e$ for some suitable edge $e$ whose tangent vector
is $\dot{e}^a$ and whose parameter length is $\epsilon$. The right
hand side is a good approximation for small $\epsilon$ or small
$A_a^i$, which allows one to expand the exponential of the
holonomy. The left hand side is not connection dependent, despite its
appearance, since the derivative in the Poisson bracket removes
$A_a^i$. For holonomies $h_e$ taking values in an Abelian group, the
right hand side is also connection independent since the two
holonomies cancel each other even after taking a derivative by
connection components. But due to non-commutativity, the cancellation
is not complete for non-Abelian holonomies. Thus, the right hand side,
which is used directly for a quantization, does become connection
dependent. As a consequence, the resulting operator will {\em not}
commute with the volume operator even though the classical expression
does not depend on momenta of the densitized triad.

This may lead to unsharp definitions of degeneracy points since it
depends on which operator is used to determine eigenstates. There must
be additional dynamical arguments to select the appropriate operator
and its eigenstates as corresponding to classical singularities. Such
situations would be much more complicated to analyze both from the
classical perspective (structure of singularities) and for quantum
dynamics.  Fortunately, every indication so far, relying on dynamical
properties of quantum gravity, points to properties as they were
anticipated from models \cite{DegFull}.

As we saw, symmetric models allow explicit investigations of many
properties expected for singularities.  It is important to keep a wide
view of all types of characteristic models since any given model by
itself may be too special as one can see it in isotropy.  Still, there
may always be properties not seen so far which may become relevant.
According to the current status, quantum hyperbolicity is realized in
much more general terms than any other mechanism to remove classical
singularities.  Moreover, it is suggested naturally by the structure
of background independent quantum gravity (e.g.\ properties of
singularities in densitized triads or the form of difference
operators). Also the non-perturbative treatment matters, as shown in
\cite{AnisoPert} where an anisotropic model was loop quantized not as
described before but as a perturbation around a loop quantized
isotropic model. Then, quantum hyperbolicity was not realized, in
contrast to the full, non-perturbative quantization of the same model.
Generalizations to less symmetric models have thus presented many
non-trivial tests of the whole framework which so far were all passed
successfully. A further test is the independence of the mechanism from
details of the matter Hamiltonian, implying that we are dealing with a
pure quantum geometry effect. Even curvature couplings, which can
arise for non-minimally coupled scalar fields, do not change the
mechanism \cite{NonMin} although at first the classical structure
seems to be quite different from that in the absence of curvature
couplings. The deep quantum picture crucially relies on spatial
discreteness and dynamical equations for a wave function. This
prevents detailed intuitive formulations, which in semiclassical
regimes are sometimes available based on effective theory. Under
certain conditions this can be applied to the transition through
classical singularities, resulting in bounce pictures to be discussed
in the next, final section.

\section{Semiclassical Pictures}

As the general scheme of quantum hyperbolicity turns out to be
realized in many cases in loop quantum gravity without counterexamples
so far, it becomes possible to ask more detailed questions about the
transition through classical singularities and in particular as to
what happened before the big bang. The general mechanism utilizes in
an essential way the discreteness of spatial geometry realized by the
loop quantization: difference equations for wave functions on the
space of triads need to be considered.

Such equations are often difficult to analyze or to solve explicitly
(see \cite{GenFunc,ContFrac,GenFuncBI,PreClassBI,InstabLRS} for some
techniques), and especially difficult to interpret. Moreover, at such
a fundamental level, the issue of how space-time emerges from the
underlying quantum state becomes exceedingly difficult especially on
inhomogeneous models with many degrees of freedom. This is an issue
faced by any quantum theory of gravity, independently of the precise
methods used. For a recent discussion in string theory, using the
AdS/CFT correspondence, see \cite{AdSCFTEmergence}.  Intuitive
pictures then usually require the use of special models allowing exact
solutions, which imposes more than just a class of symmetries such as
also specific matter ingredients (a free, massless scalar, say).
While such models can often be analyzed in great detail, their
properties may or may not be typical for general cases even within the
same class of symmetries.  This is analogous to the classical
situation where the simplest and best known cases of singularities
(isotropic cosmology, Schwarzschild black hole) may not display the
general behavior of classical singularities as they are shown to arise
from singularity theorems are from more general studies.
Nevertheless, since these types of singularities are quite relevant
for our understanding of the universe, detailed pictures for how they
can be resolved are valuable. Moreover, they can provide a basis for
more general scenarios, for instance in a perturbation theory.  The
most common intuitive picture for such resolutions is a
bounce.\footnote{Some examples for bounces from loop effects can be
  found in
  \cite{BounceClosed,Cyclic,GenericBounce,KasnerBounce,QuantumBigBang,NegCurv}.
  Oscillatory scenarios have been developed and described in
  \cite{Oscill,InflOsc,EmergentLoop,EmergentNat,LoopFluid}.}

\subsection{Effective equations}

Much intuition in cosmology derives from isotropic or at least
homogeneous models where, from a semiclassical perspective, a bounce
is the only way to resolve a classical singularity. There are only
finitely many (at most three) metric components in those models which
must stay away from zero to avoid a singularity. Thus, within a finite
amount of time all components evolve through a minimum if they were
initially contracting and thus lead to a bounce in volume. Note that
this argument makes use of the finite dimensionality of the classical
phase space and does thus not apply to inhomogeneous models. Moreover,
the argument presupposes that some kind of semiclassical description
at least of the quantum gravity state (if not matter fields) is valid
throughout, which can then be identified with a smooth spatial
geometry subject to an altered, bouncing dynamics.

While these assumptions are difficult to justify in general, or to be
avoided in bounce arguments, bounce models can be useful to study
semiclassical effects of quantum gravity. Moreover, if sufficiently
well developed, they can be used to understand how cosmological
inhomogeneities could evolve through a bounce with potential effects
on structure formation.  Rather surprisingly, one particular bounce
model provides the zeroth order basis (``free theory'') for an
effective theory of loop quantum gravity and a corresponding
perturbation scheme. The technical details are fully analogous to the
widely used low energy effective actions in particle physics and thus
very promising for developing a detailed understanding of the
semiclassical properties of loop quantum gravity.

\subsubsection{Low energy effective action}

It is often most useful to describe quantum effects in certain regimes
by correction terms to classical equations, without changing the type of
differential equations (except for possible higher derivative terms). Thus,
a quantum mechanical system would still be formulated in terms of
coupled ordinary differential equations in time rather than a partial
differential equation. Or, a quantum field theory would be described
by partial differential equations in space-time coordinates rather
than functional differential equations. This does not only lead to
technical simplifications in solution procedures but also alleviates
most interpretational issues of quantum theories.

In quantum mechanics, it is well-known that all harmonic oscillator
states follow classical trajectories exactly. There is thus no need
for modifications to classical equations of motion, and it turns out
that only a zero point energy is added to the Hamiltonian
$H_Q=\frac{1}{2m}p^2+\frac{1}{2}m^2\omega^2q^2+\frac{1}{2}\hbar\omega$.
This is similar for free quantum field theories although the zero-point
energy may diverge.  For an-harmonic contributions $U$ to the
potential or for interactions, there are corrections to the classical
Hamiltonian which can be computed in a perturbation theory around the
harmonic oscillator or a free theory. By the usual particle physics
techniques based on Legendre transforms of generating functionals of
$n$-point functions this leads to the low-energy effective action
\begin{equation} \label{GammaEff}
 \Gamma_{{\rm eff}}[q] = \int
\md t\left(\left(m+\frac{\hbar(U''')^2}
{32m^2\left(\omega^2+\frac{U''}{m}\right)^{\frac{5}{2}}}\right)\frac{\dot
q^2}{2}
-\frac{1}{2}m^2\omega^2q^2-U-\frac{\hbar\omega}{2} 
\left(1+\frac{U''}{m\omega^2}
\right)^{\frac{1}{2}}\right)
\end{equation}
showing first order corrections in $\hbar$ to the kinetic as well as
potential term of the classical action.

These low energy effective actions as they result from perturbations
around a vacuum state need to be generalized for quantum gravity. In
that case, the Hamiltonian is not only unbounded from below but also
does not allow a general notion of energy to supply meaning to ``low
energy effective action.'' (For quantizations of gravity on a
background manifold one can and often does employ low energy effective
actions. While this is valuable for scattering effects, it is not
suitable and can be highly misleading in the context of cosmology.)
Such generalizations are available and can be derived, for instance,
from a geometrical formulation of quantum mechanics: One can view the
Hilbert space of a quantum system as an infinite-dimensional phase
space whose Poisson brackets are given through the imaginary part of
the inner product. This satisfies all mathematical conditions imposed
in the definition of Poisson brackets.

To derive the Poisson structure in a simple way, we choose expansion
coefficients in states $|\psi\rangle=\sum_jc_j|\psi_j\rangle$ with
respect to an arbitrary basis $\{|\psi_j\rangle\}$ as coordinates on
the Hilbert space. Then, the imaginary part of the inner product of
states can be interpreted as Poisson structure
\[
 \{{\rm Re} c_j,{\rm Im} c_k\}= \frac{1}{2\hbar}\delta_{jk}
\]
equipping the Hilbert space with a phase space structure.
Furthermore, the Schr\"odinger equation for $|\psi\rangle$ turns out
to be equivalent to Hamiltonian equations of motion generated by the
quantum Hamiltonian $H_Q(\{c_j\})=\langle\hat{H}\rangle$. Here, $H_Q$
is a function on the phase space with coordinates $c_j$ by using the
state with expansion coefficients $c_j$ to compute the expectation
value of $\hat{H}$.  That this gives the correct dynamics follows
easily by choosing the eigenbasis $\{|\psi_j\rangle\}$ of $\hat{H}$.
Then, $H_Q(\{c_j\})=\sum_{i,j}
\bar{c}_ic_j\langle\psi_i|\hat{H}|\psi_j\rangle= \sum_jE_j|c_j|^2$
with energy eigenvalues $E_j$, $\hat{H}|\psi_j\rangle=
E_j|\psi_j\rangle$, and the Hamiltonian equations of motion are
\begin{equation}
 \frac{\md}{\md t}{\rm Re} c_j = \{{\rm Re} c_j,H_Q\}=
 \frac{E_j}{\hbar}{\rm Im} c_j \quad,\quad
\frac{\md}{\md t}{\rm Im} c_j =
\{{\rm Im} c_j,H_Q\}= -\frac{E_j}{\hbar}{\rm Re} c_j\,.
\end{equation}
This is equivalent to $\dot{c}_j=-i\hbar^{-1}E_j c_j$ in agreement
with the Schr\"odinger evolution.  Quantum mechanics is thus formally
much closer to classical mechanics, making a relation through
effective equations possible in suitable approximations. While the
full quantum dynamics already appears in classical form for the
infinitely many variables $c_j$, any effective approximation must
include a truncation to finitely many variables for a mechanical
system.

\subsubsection{Quantum variables}

A more useful set of coordinates for this purpose is defined as
follows: we use ``classical'' variables\footnote{This term often gives
  rise to confusion as these variables are close to the classical ones
  (e.g.\ in the sense of describing the peak position of a single wave
  packet) only in semiclassical regimes while they are defined here
  for an arbitrary quantum system. The justification for the term is
  that irrespective of the regime the quantum phase space is a fiber
  bundle over the classical phase space with bundle projection
  $|\psi\rangle\mapsto
  (\langle\psi|\hat{q}|\psi\rangle,\langle\psi|\hat{p}|\psi\rangle)$
  \cite{Schilling}.}
\begin{equation}
q=\langle\hat{q}\rangle \quad\mbox{ and }\quad
  p=\langle\hat{p}\rangle
\end{equation}
and quantum variables 
\begin{equation} \label{QuantVar}
 G^{a,n}:=\langle(\hat{q}-\langle\hat{q}\rangle)^{n-a}
(\hat{p}-\langle\hat{p}\rangle)^a\rangle_{\rm Weyl}
\end{equation}
where $n\geq 2$, $a=0,\ldots,n$ and ``Weyl'' denotes the totally
symmetric ordering of the operators before the expectation value is
taken.  Poisson relations of these variables are related to
commutators as we will see in more detail later:
$\{q,p\}=\langle[\hat{q},\hat{p}]\rangle/i\hbar=1$ and
$\{q,G^{a,n}\}=0=\{p,G^{a,n}\}$. For $\{G^{a,n},G^{b,m}\}$ a closed
formula exists but is rather lengthy \cite{EffAc,EffectiveEOM}.

Quantum variables are dynamical just as the classical variables are:
for a semiclassical state they change in time, e.g.\ if a wave packet spreads
and deforms. The resulting evolution back-reacts on the classical
variables which determine the peak position of the wave packet. Their
motion then in general differs from the classical one, which is to be
captured in appropriate quantum correction terms to the classical
equations of motion.  The exact behavior is determined by the
Schr\"odinger equation, or equivalently by the quantum Hamiltonian
$H_Q$ which couples classical and quantum variables. For instance, for
a cubic potential we have $\langle\hat{q}^3\rangle=
q^3+6qG^{0,2}+6G^{0,3}$ with a coupling term $qG^{0,2}$ in addition to
a zero-point contribution by $G^{0,3}$.

More generally, for an an-harmonic oscillator with classical Hamiltonian
$H=\frac{1}{2m}p^2+\frac{1}{2}m\omega^2q^2+U(q)$ we have
a quantum Hamiltonian with infinitely many coupling terms,
\begin{eqnarray}
  H_Q &=& \langle H(\hat{q},\hat{p})\rangle=\langle
  H(q+(\hat{q}-q),p+(\hat{p}-p))\rangle\\
  &=&\frac{1}{2m}p^2+\frac{1}{2}m\omega^2q^2+U(q)
  +\frac{\hbar\omega}{2}(\tilde{G}^{0,2}+\tilde{G}^{2,2})
  +\sum_{n>2}\frac{1}{n!}
  \left(\frac{\hbar}{m\omega}\right)^{n/2}U^{(n)}(q)\tilde{G}^{0,n} 
 \label{HQexpand}
\end{eqnarray}
written in dimensionless variables $\tilde
G^{a,n}=\hbar^{-n/2}(m\omega)^{n/2-a}G^{a,n}$. As indicated, these
terms follow from formally expanding the Hamiltonian in $\hat{q}-q$
and $\hat{p}-p$.
Using the Poisson brackets of all classical and quantum variables,
$H_Q$ generates Hamiltonian equations of motion $\dot{f}=\{f,H_Q\}$:
\begin{eqnarray}
\dot{q}&=& \frac{p}{m}\label{eom}\\
\dot{p}&=&-m\omega^2q -U'(q)-\sum_n\frac{1}{n!}\left(
\frac{\hbar}{m\omega}\right)^{n/2}U^{(n+1)}(q)\tilde{G}^{0,n}\\
\dot{\tilde{G}}{}^{a,n}&=&-a\omega
\tilde{G}^{a-1,n}+(n-a)\omega \tilde{G}^{a+1,n}
-a\frac{U''(q)}{m\omega}\tilde{G}^{a-1,n}\\
\nonumber&&+
\frac{\sqrt{\hbar}aU'''(q)}{2(m\omega)^{\frac{3}{2}}}\tilde{G}^{a-1,n-1}
\tilde{G}^{0,2}
+\frac{\hbar aU^{''''}(q)}{3!(m\omega)^2}\tilde{G}^{a-1,n-1}\tilde{G}^{0,3}\\
\nonumber&&
-\frac{a}{2}\left(
\frac{\sqrt{\hbar}U'''(q)}{(m\omega)^{\frac{3}{2}}}
\tilde{G}^{a-1,n+1}+\frac{\hbar
U^{''''}(q)}{3(m\omega)^2}\tilde{G}^{a-1,n+2}\right)+\cdots\,.
\end{eqnarray}
These are infinitely many coupled equations for infinitely many
variables. At this stage, the system is still fully equivalent to the
Schr\"odinger dynamics, just written in terms of ``$n$-point
functions'' $q$, $p$ and $G^{a,n}$ instead of the wave function they
determine.

As an example we can look at the harmonic oscillator whose Hamiltonian
equations of motion are
\begin{eqnarray*}
\dot{p}=&\{p,H_Q\}&=-m\omega^2 q\\
\dot{q}=&\{q,H_Q\}&=\frac{1}{m} p\\
\dot{G}^{a,n}=&\{{G}^{a,n},H_Q\}&=\frac{1}{m}(n-a)
G^{a+1,n}-m\omega^2
aG^{a-1,n}\,.
\end{eqnarray*}
In this case, all terms coupling the $G^{a,n}$ for different $n$
vanish, and we have an infinite set of differential equations only
finitely many of which are coupled to each other. Moreover, the
equations are linear and can easily be solved.  For instance, constant
solutions for the quantum variables exist, satisfying uncertainty
relations such as $G^{0,2}G^{2,2}\geq\frac{\hbar^2}{4}+(G^{1,2})^2$.
These solutions correspond to the well-known coherent states which do
not spread and follow the classical trajectories exactly.  But even if
quantum variables are not constant, they do not appear in equations of
motion for the classical variables and thus do not back-react on them.
This is why the effective action of the harmonic oscillator is
identical to its classical action.

With an-harmonic contributions, coupling terms are switched on and all
equations get coupled to each other. Consistent truncations to
finitely many equations for finitely many variables are then required
for an effective approximation. This is possible in, e.g., an
adiabatic approximation \cite{EffAc}: we solve approximately for the
leading $G^{a,n}$ assuming that they change in time much more slowly
than $q$ and $p$, and insert the solutions into the equations of
motion for $q$ and $p$. Doing this to first order in $\hbar$ and to
second order in the adiabatic approximation, and writing the first
order equations for $q$ and $p$ as a second order equation for $q$, we
obtain
\begin{eqnarray*}
&&\nonumber\left(m+\frac{\hbar U'''(q)^2}{32m^2\omega^5\left(
 1+\frac{U''(q)}{m\omega^2}\right)^{5/2}}\right)\ddot q
+\frac{\hbar\dot
 q^2\left(4m\omega^2U'''(q)U''''(q)\left(1+\frac{U''(q)}{m\omega^2}\right)-
 5U'''(q)^3\right)}
 {128m^3\omega^7\left(1+\frac{U''(q)}{m\omega^2}\right)^{7/2}}\\
&&+m\omega^2q+U'(q)+\frac{\hbar
U'''(q)}{4m\omega\left(1+\frac{U''(q)}{m\omega^2}\right)^{1/2}}=0
\end{eqnarray*}
as it also follows from the low energy effective action (\ref{GammaEff}).

\subsubsection{Solvable systems}

The derivation clearly shows the role of the harmonic oscillator: its
classical and quantum variables decouple to sets of finitely many
linear equations and there is no back-reaction by quantum variables on
expectation values.  This happens whenever the Hamiltonian is
quadratic in canonical variables, or more generally when the system is
linear, i.e.\ when the Hamiltonian and a set of basic variables form a
Lie algebra using Poisson brackets. Effective equations can then be
obtained by perturbing around the exact solutions of such a solvable
system.

Realizing this opens one possibility for generalizations of low energy
effective equations, which are then based on alternative solvable
systems suitable for a given context. Remarkably, examples for such
solvable models are realized in cosmology and allow a systematic
effective theory in this context.

\subsection{Large scale effective theory for cosmological bounces}

Identifying such a solvable model suitable for cosmology and deriving
its perturbation equations allows one to derive intuitive
semiclassical pictures which describe the transition through classical
singularities in detail.

\subsubsection{Solvable model for cosmology}

The harmonic oscillator with its periodic motion is not suitable as a
solvable system for cosmology. But we can look at a free isotropic
scalar model whose Friedmann equation $c^2\sqrt{p}=\frac{4\pi
  G}{3}p^{-3/2}p_{\varphi}^2$ follows from (\ref{IsoHam}) with
variables $c=\dot{a}$ (extrinsic curvature) and $p^{3/2}=a^3$ (volume)
where $p$ is the densitized triad component, now assumed positive.
Solving for $p_{\varphi}$ yields $p_{\varphi}\propto cp=:H$, to be
interpreted as the Hamiltonian which generates the flow in the
variable $\varphi$ playing the role of internal time.  This
Hamiltonian is quadratic,\footnote{We were not careful about the signs
  involved when solving the quadratic constraint equations for
  $p_{\phi}$, although one can conclude from the constraint only that
  $|p_{\phi}|\propto|cp|$. Using the absolute value for $H$, on the
  other hand, would not leave it strictly quadratic. As we are mainly
  interested in states for which the expectation value of $\hat{H}$ is
  large compared to the spread $\Delta H$, we do not need to worry
  about significant contributions from solutions with different signs
  of the Hamiltonian.  Note that $H$ and $\Delta H$ are preserved in
  time. Thus, if the condition $H\gg\Delta H$ is satisfied once, e.g.\
  for an initial semiclassical state, it will be satisfied at all
  values of $\phi$.}  although not of the harmonic oscillator form.
But as in this case, classical and quantum variables decouple and the
quantum Hamiltonian $H_Q=cp+G^{cp}$ is obtained, as in
(\ref{HQexpand}), by adding only the zero point contribution
$G^{cp}=\frac{1}{2}(\langle \hat{c}\hat{p}\rangle+
\langle\hat{p}\hat{c}\rangle)-cp$. (Here, we use a slightly different
notation compared to the general quantum variables (\ref{QuantVar})
for better clarity.)

Its equations of motion follow by using Poisson brackets
$\{c,G^{cp}\}=0= \{p,G^{cp}\}$ as well as $\{G^{cc},G^{cp}\} =
2G^{cc}$, $\{G^{cc},G^{pp}\} = 4G^{cp}$, $\{G^{cp},G^{pp}\} =
2G^{pp}$ and further ones depending on which variables one is
interested in solving for.
We will now show how the relevant Poisson relations are derived, using
the example of $\{G^{0,2},G^{1,2}\}$. The basic identity is the relation
\begin{equation} \label{PoissonComm}
 \{\langle\hat{A}\rangle,\langle\hat{B}\rangle\}=
 \langle[\hat{A},\hat{B}]\rangle/i\hbar
\end{equation}
which clearly shows how the quantum Poisson brackets are related to
commutators. The familiar relation
$\{q,p\}=\langle[\hat{q},\hat{p}]\rangle/i\hbar=1$ then follows
immediately. But it cannot be directly applied to quantum variables
such as $G^{cc}=\langle\hat{c}^2\rangle-c^2$ and
$G^{cp}=\frac{1}{2}(\langle \hat{c}\hat{p}\rangle+
\langle\hat{p}\hat{c}\rangle)-cp$ since here also products of
expectation values occur. This can easily be dealt with using the
Leibniz rule
\begin{equation}
 \{f,g_1g_2\}=g_1\{f,g_2\}+ \{f,g_1\}g_2
\end{equation}
to reduce all Poisson brackets to those of expectation values of
operators where (\ref{PoissonComm}) applies.
With the brackets $\{\langle
\hat{c}^2\rangle, \langle\hat{c}\hat{p}\rangle\}=
2\langle\hat{c}^2\rangle$, $\{\langle
\hat{c}^2\rangle,cp\}= 
\{\langle\hat{c}^2\rangle,\langle\hat{c}\rangle\langle\hat{p}\rangle\}=
c\{\langle \hat{c}^2\rangle,\langle\hat{p}\rangle\}=2c^2$ and
$\{c^2,cp\}=2c^2$ we then derive
\[
\{G^{cc},G^{cp}\}=\{\langle\hat{c}^2\rangle-c^2,
 {\textstyle\frac{1}{2}}(\langle
\hat{c}\hat{p}\rangle+ \langle\hat{p}\hat{c}\rangle)-cp\}=
2(\langle\hat{c}^2\rangle-c^2)=2G^{cc}\,.
\]
Similarly, all other Poisson brackets are derived, for which also
closed formulas exist \cite{EffAc,EffectiveEOM}.

Equations of motion generated by $H_Q=cp+G^{cp}$ are thus
$\dot{c}=c$, $\dot{p}=-p$ for the classical variables, and
$\dot{G}^{cc}=2G^{cc}$, $\dot{G}^{cp}=0$ and
$\dot{G}^{pp}=-2G^{pp}$ for the quantum variables. This is easily
solved by $c(t)=c_1e^{t}$, $p(t)=c_2e^{-t}$, $G^{cc}(t)=c_3e^{2t}$,
$G^{cp}(t)=c_4$ and $G^{pp}(t)=c_5e^{-2t}$ with suitable integration
constants which are only restricted by the uncertainty relation
$c_3c_5\geq\hbar^2/4+c_4^2$. Although constant solutions of the
quantum variables do not exist, the semiclassical properties are quite
similar to those of the harmonic oscillator. In particular,
semiclassicality is preserved: the relative spreads $\Delta
p/p=\sqrt{G^{pp}}/p$ and $\Delta c/c=\sqrt{G^{cc}}/c$ are constant
throughout the whole evolution. This will allow us to derive effective
equations and to develop a perturbation theory around the known
solutions of the solvable model where the free isotropic scalar plays
the same role as the harmonic oscillator in quantum mechanics or free
theories in quantum field theory. It also shows that conclusions drawn
from the free isotropic scalar have to be taken with great care, just
as the behavior of the harmonic oscillator is not at all typical for
general quantum systems even in semiclassical regimes.

\subsubsection{Loop formulation and bounces}

In a loop formulation we do not use the Schr\"odinger quantization of
basic variables $c$ and $p$. Instead, in a loop quantization
(\ref{IsoRep}) the operator $\hat{p}$ has a discrete spectrum and no
operator for $c$ exists. What is represented are only exponentials
$\exp(ic)$ such that, e.g., $\sin c$ occurs instead of $c$. The
resulting Hamiltonian operator
$\hat{H}=-\frac{1}{2}i(\exp(ic)-\exp(-ic))\hat{p}$ is a shift operator
and implies a difference equation for the state in a triad
representation. This operator is not identical to (\ref{IsoHamOp})
derived earlier but is closely related. What we have not included here
are quantum effects in the inverse power $p^{-3/2}$ of the matter
Hamiltonian as they occur in (\ref{da}). Such terms do not allow
solvable models but can be included in perturbation theory.

The Hamiltonian is non-quadratic and not solvable in an obvious way.
But introducing $\hat{J}=\hat{p}\widehat{e^{ic}}$ allows us to reorder
the Hamiltonian to become a linear expression
$\hat{H}=-\frac{1}{2}i(\hat{J}-\hat{J}^{\dagger})$. The price to pay
is that the algebra of basic operators $\hat{p}$ and $\hat{J}$ is
non-canonical, which usually implies that the system is not solvable
in the above sense even for a linear Hamiltonian. But for the system
under consideration it turns out that the set of Hamiltonian operator
and basic variables $(\hat{p},\hat{J})$ forms a linear system
\cite{BouncePert}, given by the (trivially) centrally extended ${\rm
  sl}(2,{\mathbb R})$ algebra
\begin{equation}
 [\hat{p},\hat{J}]=\hbar\hat{J}\quad,\quad
 [\hat{p},\hat{J}^{\dagger}]=-\hbar\hat{J}^{\dagger}\quad,\quad{}
 [\hat{J},\hat{J}^{\dagger}]=-2\hbar\hat{p}-\hbar^2\,.{}
\end{equation}
Taking expectation values, the linear quantum Hamiltonian
$H_Q=-\frac{1}{2}i(J-\bar{J})$ does not even receive a zero point
contribution and generates equations of motion
\begin{equation}
 \dot{p}=\{p,H_Q\}=-{\textstyle\frac{1}{2}}(J+\bar{J})\quad,\quad
\dot{J}=\{J,H_Q\}=-{\textstyle\frac{1}{2}}(p+\hbar)=\dot{\bar{J}}
\end{equation}
with solution
\begin{equation}
 p(t) = {\textstyle\frac{1}{2}}(c_1e^{-t}+c_2e^{t})-
 {\textstyle\frac{1}{2}}\hbar \quad,\quad
 J(t) = {\textstyle\frac{1}{2}}(c_1e^{-t}-c_2e^{t})+iH\,.
\end{equation}
They are simply linear combinations of the solutions in the
Schr\"odinger quantization. Depending on whether $c_1c_2$ is positive
or negative we have bouncing solutions of $\cosh$ form or solutions of
$\sinh$ form which arrive at the classical singularity $p=0$ after a
finite amount of internal time $t$. For our purpose of studying
singularity removal we have to analyze these types of solutions in
more detail.

But we have not yet implemented all necessary conditions and some of
those solutions are non-physical. Classically we have $J\bar{J}=p^2$
for $J=p\exp(ic)$, which as a reality condition for $c$ must have an
analog in the quantization.  For states, this corresponds to
normalizability in a physical inner product ensuring that $\exp(ic)$
is quantized to a unitary operator. Although computing a physical
inner product is usually a difficult issue, for the solutions $p(t)$
and $J(t)$ of expectation values we can implement the reality
condition directly by just noticing that $J\bar{J}=p^2+O(\hbar)$ must
remain satisfied at the effective level.  Only corrections of order
$\hbar$ to the classical condition may result since
$\langle\hat{J}\rangle \langle\hat{J}^{\dagger}\rangle\not=\langle
\hat{J}\hat{J}^{\dagger}\rangle$.  The only way to implement this at
all times is by requiring
\begin{equation}
 \frac{J\bar{J}}{(p+\hbar/2)^2}= \frac{(c_1e^{-t}-c_2e^{t})^2+
4H^2}{(c_1e^{-t}+c_2e^{t})^2}=1
\end{equation}
which implies $c_1c_2=H^2$ up to quantum corrections which are not
important for large $H$.
This leaves only the bouncing solution
\begin{equation}
 p(t)=H\cosh(t-\delta)-\hbar\quad,\quad
J(t)=-H(\sinh(t-\delta)+i)
\end{equation}
with a single constant of integration $\delta$. Note that the minimum
of $p$ is given by $H-\hbar$ which for a large Hamiltonian $H$, i.e.\
large matter content, is far away from the classical singularity at
$p=0$. The bounce trajectory agrees well with numerical solutions of
(physically normalized) wave packets in a closely related model
studied recently in \cite{QuantumBigBang,APS}.

\begin{figure}
  \includegraphics[height=.25\textheight]{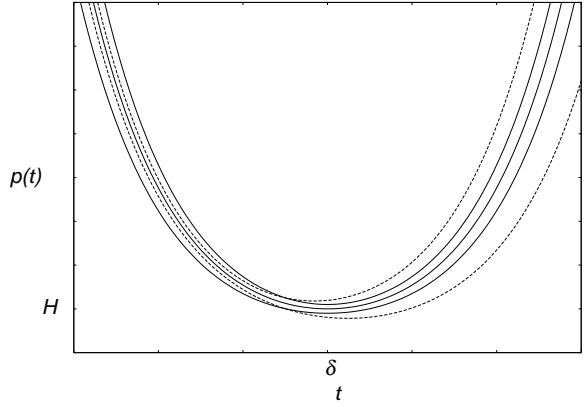}
  \caption{Bouncing effective solutions for expectation value and spread.
    \label{EffBounce}}
\end{figure}

Similarly, we can compute equations of motion for the spread parameters
\begin{equation}
\dot{G}^{pp}=-2G^{pJ}\quad ,\quad\dot{G}^{JJ}=-2G^{pJ}
\quad,\quad \dot{G}^{pJ}=
-{\textstyle\frac{1}{2}}G^{JJ}-{\textstyle\frac{3}{2}} G^{pp}-
{\textstyle\frac{1}{2}}(p^2-J\bar{J}+\hbar p+\hbar^2/2)
\end{equation}
(using $\hat{J}\hat{J}^{\dagger}=\hat{p}^2$ and the commutation
relations).  They satisfy the uncertainty relation
\[
 G^{pp}G^{JJ}-|G^{pJ}|^2\geq\frac{\hbar^2}{4}|J|^2\,.
\]
For $H\gg\hbar$, a solution is given by $(\Delta p)^2=G^{pp}\approx
\hbar H\cosh(2(t-\delta_2))$. The semiclassical behavior throughout
the bounce is clearly seen from Fig.~\ref{EffBounce} where during the
contraction and expansion phases the relative spreads are almost
constant as in the Schr\"odinger quantization, although they may
change from the contracting to the expanding phase depending on the
integration constant $\delta-\delta_2$. The loop quantization thus
connects the two branches in a well-defined way but leaves open the
relative degrees of semiclassicality on both sides.

\subsubsection{Perturbations}

As described in general terms before, perturbations around solvable
models can be formulated by standard means. This allows one to
introduce terms which imply couplings between classical and quantum
variables, such as matter potentials, inhomogeneous degrees of
freedom, or basic issues of a loop quantization. The latter effects
include different factor orderings of the constraint or modifications
due to inverse powers in the matter Hamiltonian. This provides a
systematic way to derive the evolution of fields through cosmological
bounces, which is not available in other schemes where assumptions
about the regularity at a bounce must be imposed rather than being
derivable. The perturbation scheme mentioned here can certainly break
down, which implies that the semiclassical nature of realistic bounces
is testable in a self-consistent manner.

In addition to conceptual lessons, mechanisms for singularity
resolution provide a possible relation to observations in bounce
scenarios of structure formation, making use of Hamiltonian
perturbation theory \cite{HamPerturb} and an implementation in
perturbative loop quantum gravity \cite{QuantCorrPert}.

\section{The current status of singularity resolution in quantum gravity}

We have focused the discussions here on the singularity problem from a
general perspective. Accordingly, detailed scenarios we described were
only those which have been formulated in a sufficiently general
context, applying to more than a small class of situations and not
showing any obvious limiting assumptions.

With recent developments, loop quantum cosmology has provided the
first systematic effective theory for the non-singular evolution of
perturbations through a classical singularity using perturbation
theory around bouncing solutions.  Expectation values and the
spreading of semiclassical states can be computed explicitly.  This
shows under which circumstances semiclassical states are obtained also
during and after the bounce, although the degree of semiclassicality
depends on initial conditions.  A general analysis is unfinished. In
general, bounce pictures are difficult to extend to inhomogeneities
unless the latter remain perturbative. Often, one appeals to the BKL
conjecture to suggest that homogeneous scenarios might be sufficient
to discuss singularity resolution. But by construction bounces in
general do not allow one to go sufficiently deep in the regime,
asymptotically close to a classical singularity, which is used in the
BKL scenario to argue that time derivatives dominate spatial
derivatives.

Under unrestricted perturbations, bounce pictures can easily break
down and full quantum properties are required.  Even in this case,
singularity removal in loop quantum gravity is established in many
situations by quantum hyperbolicity: wave functions on spaces of
triads can be extended uniquely across classical singularities.  This
crucially rests on quantum geometry (rather than matter) and
especially its discrete spatial structure.  Background independence is
important for the detailed realization, providing many consistency
tests for quantum gravity. Quantum hyperbolicity is at present the
most general mechanism for non-singular behavior, indeed showing
clearly the role of important aspects of a background independent
quantization of gravity.  Nevertheless, much remains to be understood
for a general statement on singularity resolution. The main problem at
the current stage of developments is not only the complexity of
quantum gravity, but also an understanding of the classical structure
of singularities needed to select, in the absence of general
solutions, the quantum regime to be looked at.

\begin{theacknowledgments}
  The author is grateful to Mario Novello for an invitation to give a
  lecture course at the XIIth Brazilian School on Cosmology and
  Gravitation on which this article is based. He also profitted from
  talks and discussions presented at the workshop ``The Quantum Nature
  of Spacetime Singularities'' held at the KITP, Santa Barbara, in
  January 2007, and thanks Claus Kiefer for comments on possible
  singularity removal in a Wheeler--DeWitt quantization. This research
  was supported in part by the National Science Foundation under Grant
  Nos.\ PHY99-07949 and PHY05-54771.
\end{theacknowledgments}

\end{document}